\pdfoutput=1
\documentclass[aps,pra,reprint,showpacs]{revtex4-1}
\usepackage{graphicx}
\usepackage{amsmath}
\usepackage{amssymb}
\usepackage{amsfonts}
\usepackage{dcolumn}
\usepackage{dsfont}
\usepackage{latexsym}
\usepackage{rotating}
\usepackage{color}
\usepackage{latexsym}
\usepackage{bbm}
\usepackage{subfigure}
\usepackage{float}
\usepackage{epsfig}
\usepackage{psfrag}
\usepackage{natbib}
\usepackage{bm}
\usepackage{amsthm}
\usepackage{eucal}
\usepackage{mathrsfs}
\usepackage{url}
\usepackage{braket}
\usepackage{mathtools}
\usepackage{amsmath,amssymb}
\usepackage{epsfig}
\usepackage{amsmath,amssymb}
\usepackage{hyperref}

\usepackage{color}

\usepackage{hyperref}
\hypersetup{
colorlinks=true,final=true,
        linkcolor=blue,
        citecolor=blue,
        filecolor=blue,
        urlcolor=blue,
}

\begin{document}

\title{Collision-mediated radiation due to Airy-soliton interaction in nonlinear Kerr medium}

\author{Aritra Banerjee$^\star$ and Samudra Roy$^\dagger$}
\affiliation{Department of Physics, Indian Institute of Technology Kharagpur, W.B. 721302, India}

\email{$^\dagger$samudra.roy@phy.iitkgp.ernet.in\\
$^\star$aritra@iitkgp.ac.in}

\begin{abstract}

We study the interaction of a co-propagating finite energy Airy pulse and a soliton in a Kerr medium under third order dispersion. It is observed that a strong radiation appears when self-accelerating Airy pulse collides with a delayed soliton in time domain. We confirm both analytically and numerically that this radiation can only be initiated in the environment of third order dispersion (TOD). The radiation frequency is red or blue shifted depending on the numeric sign of TOD parameter.  We adopt a simple Si-based waveguide to validate the normalised parameter that we use for the analysis. We develop a theory to explain the collision-mediated radiation and our theoretical results are found to be in good agreement with the direct numerical modelling of the  generalised nonlinear Schr\"{o}dinger equation (GNLSE).         

\end{abstract}

\maketitle

 \section{Introduction}

There have been many studies about the temporal pulse interactions in linear and nonlinear optical mediums. Such interaction mechanism is useful to study because it opens up the possibility of controlling light by light. The physics of light interaction often leads to some fascinating analogies in physics ranging from hawking radiation to black hole physics to even temporal analogue of spatial reflection and refraction.  \cite{Robertson,Wang,Gorbach,Demircan,Philbin,Belgiorno,Faccio}. As a specific example, an intense optical pulse can be used to generate optical event horizon where any weak pulse can be reflected from the optical refractive index barrier generated by the intense pulse in Kerr nonlinear medium \cite{Webb}. Recently the temporal analogue of reflection and refraction phenomenon is described using the interaction of a pulse approaches a moving temporal boundary  across which  refractive index changs \cite{Agrawal}. The interaction of the optical soliton with the continuous waves and the dispersive waves has also been studied in details \cite{Biancalana,Skryabin,Yulin}. The interaction between the optical pulses in temporal domain requires the pulses to move with different relative group velocities  such that one can approach to other and interact at some point. The collision becomes more prominent and gives rise to exciting effects if one of the pulses is accelerating (or decelerating) in nature. Interaction of a Raman soliton with the linear waves is an example of such collision which occurs during  supercontinuum generation process \cite{Dmitry}. The optical pulses cannot accelerate naturally without higher order effects with the exception of the Airy pulse, a relatively new kind of optical pulse introduced to optics \cite{Berry,Siviloglou,Siviloglou_b,Broky}. These pulses are temporal analogues of Airy beams whose field envelope is described by the Airy function. An ideal Airy pulse resists the linear dispersion and maintains its shape indefinitely. However the ideal Airy waveform has infinite energy and is not square integrable. For practical realization we generally apodize the pure Airy pulse with an exponential decay function. The apodized Airy pulse still resist the dispersion effect but for a finite propagation distance.

In this work, we try to explore the collision dynamics between a finite energy Airy pulse and a co-propagating soliton in a nonlinear dispersive medium. If the soliton and the Airy pulse are launched in an optical medium with the same central frequency and an initial time offset, the Airy pulse will decelerate and will reach the soliton at a certain distance even though the  group velocities are same at the launching point. The interaction of the soliton and the Airy pulse have been reported earlier where the pulses have same or different central frequency \cite{Rudnick,Cai}. It has been shown that the Airy pulse can alter the components of a soliton much more efficiently than the other optical pulses due to its resilience to the dispersive effects of the medium. The impact of the collision between the self-accelerating Airy pulse and the soliton becomes dramatic under the effect of third order dispersion (TOD). The Airy-soliton collision (in time domain) leads to a unique kind of radiation (in frequency domain) never explored before. Unlike the traditional phase-matching radiation \cite{Skryabin}, the present radiation is collision dependent. We establish a complete theory to understand the underlying physics of such radiation.  

The paper is organized as follows. In Sec. II we introduce the mathematical structure of the problem and study the Airy-soliton collision under linear dispersion. We extend the study in Sec. III by including TOD and propose a realistic structure exhibiting ideal environment for collision mediated radiation. Further we establish a complete theory for radiations generated under positive and negative TOD. The analytical results are compared with full numerical simulations. Finally we summarize our findings in Sec IV.

 \section{Collision dynamics without higher order dispersion}
\begin{figure}[h!]
   \begin{center}
   
   \includegraphics[trim=4.6in 0.05in 5.6in 0.6in,clip=true,  width=42mm]{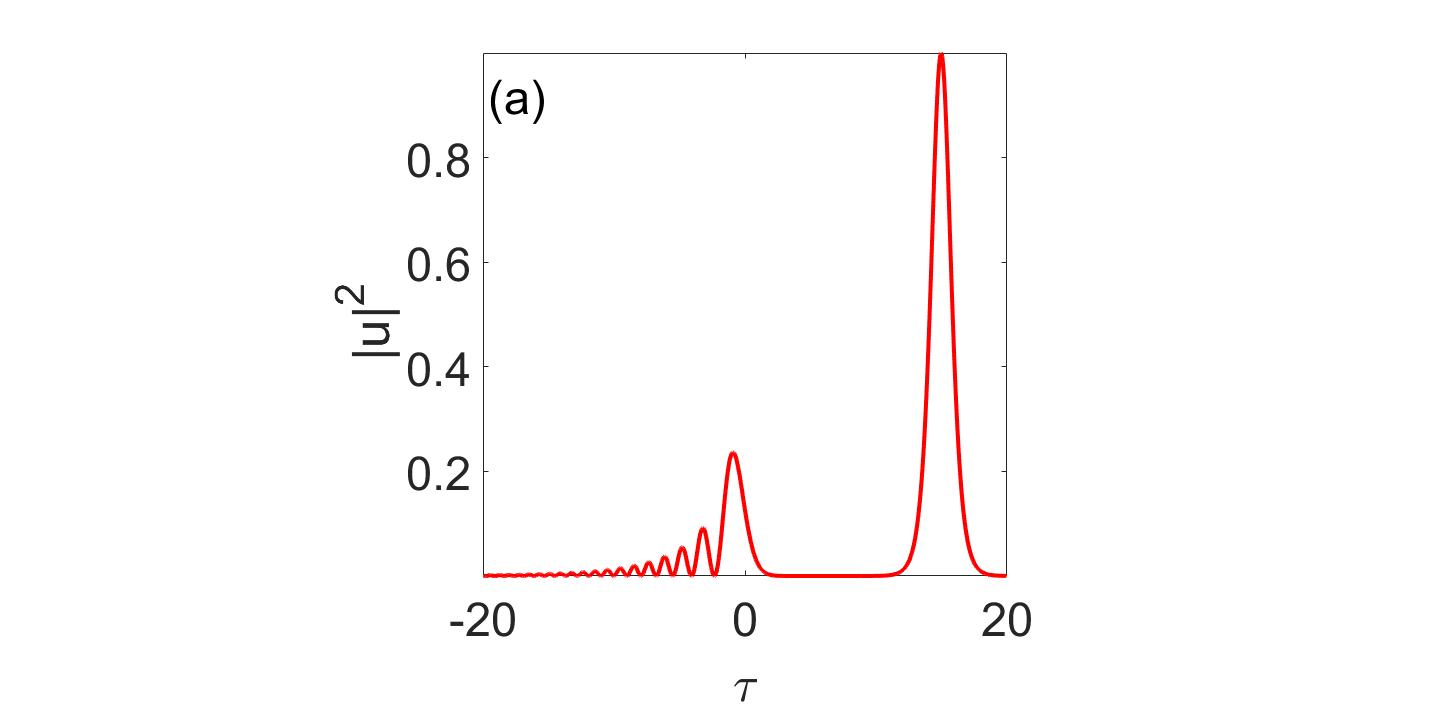}
   \includegraphics[trim=4.6in 0.05in 5.6in 0.6in,clip=true,  width=42mm]{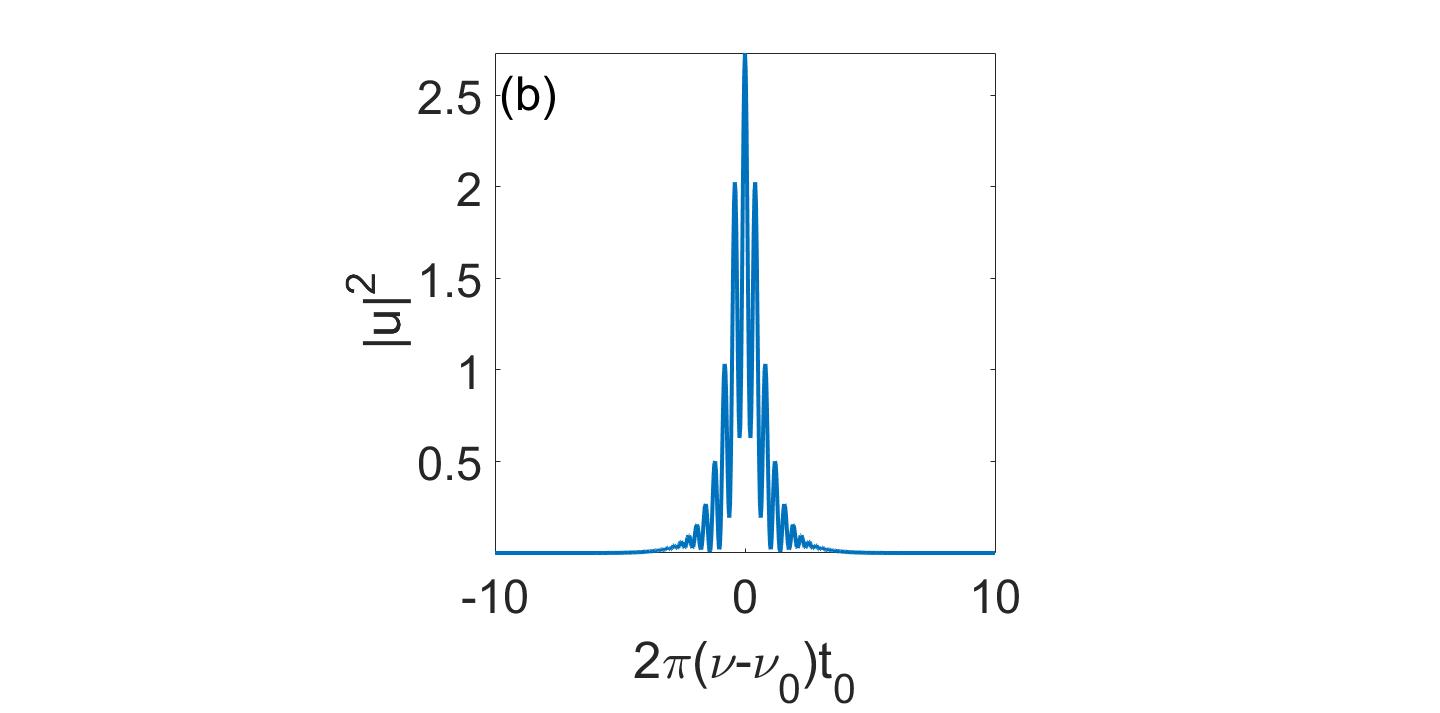}
   \includegraphics[trim=4.0in 0.05in 5.6in 0.6in,clip=true,  width=42mm]{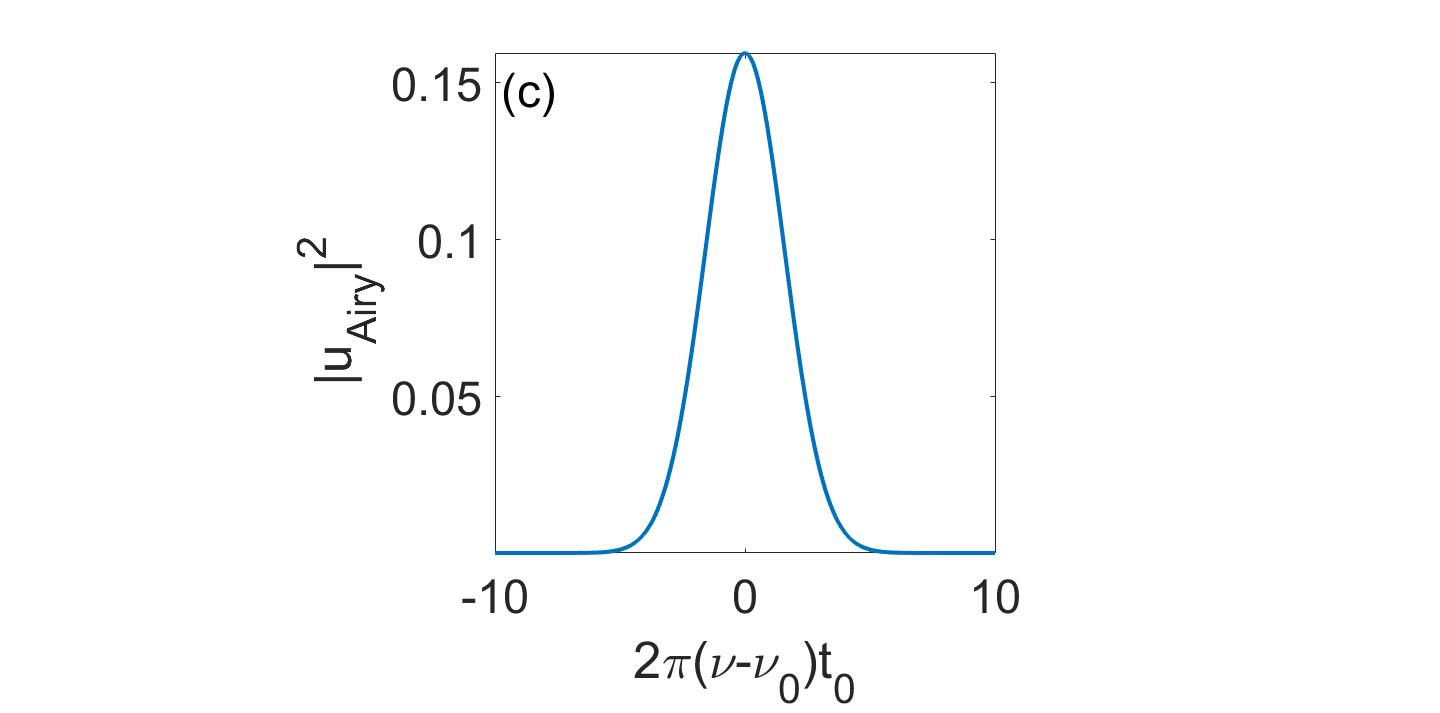}
   \includegraphics[trim=4.0in 0.05in 5.6in 0.6in,clip=true,  width=42mm]{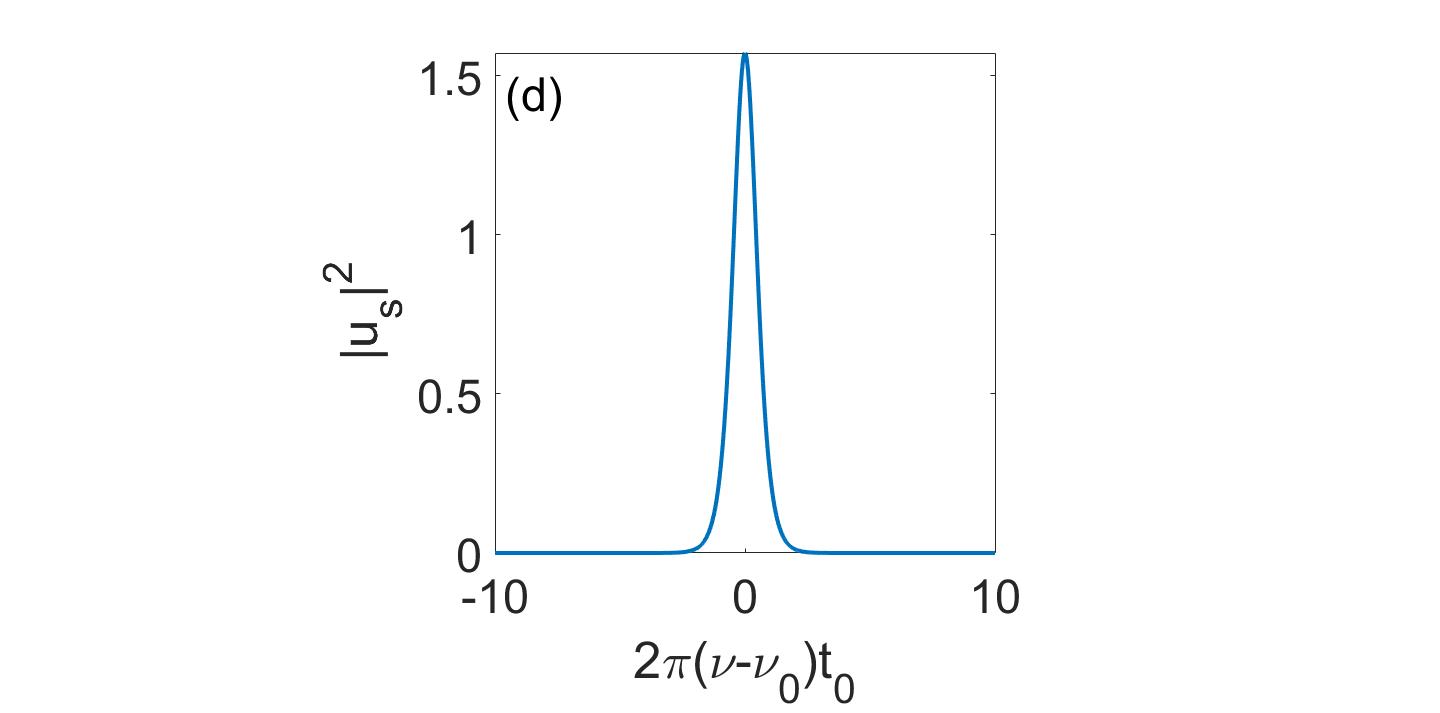}
   
   \vspace{-1em}
   \caption{(a)Temporal  and (b) spectral distribution  of the pulses at the launching point with $\Delta \tau=15$.(c) Initial  spectrum of a truncated Airy pulse (with $a=$0.1)  and (d) soliton where $u_s=sech(\tau-\Delta \tau)$.} 
   \label{fig1}
   \end{center}
   \end{figure}

\noindent In this section, we first study the interaction among Airy pulse and soliton in absence of TOD. The two pulses are launched with a time separation $\Delta \tau$. As the Airy pulse de-accelerates, it is launched earlier so that, at some spatial distance it catches up with the soliton that moves with a constant group velocity $v_g$. The dynamics of the total field is governed by well-known NLSE which has the following form in normalised unit \cite{Agarwal_book},
  \begin{equation}\label{q1}
 \partial_\xi u=-\frac{i}{2}sgn(\beta_2)\partial^2_{\tau}u+i|u|^2u          
  \end{equation}
Here $u(\xi,\tau)$ is the total field envelope and $\beta_2$ is the group velocity dispersion coefficient with unit, ps$^2$/m. The space ($\xi$) and retarded time ($\tau$) coordinate is resealed as, $\xi=zL_D^{-1}$ and $\tau=(t-z/v_g)t_0^{-1}$, where $t_0$ is the soliton width and dispersion length, $L_D=t_0^2|\beta_2|^{-1}$. We consider soliton order ($N$) as 1.   We write the initial field as,
  \begin{equation}\label{q2}
  u(0,\tau)= Ai(\tau)\exp(a\tau)+sech(\tau-\Delta \tau)
  \end{equation} 
In the above equation (Eq.\ref{q2})  $a$ is the truncation parameter  which truncates the infinite energy Airy pulse to the physically realizable finite energy pulse and $\Delta\tau$ is the initial time-delay between the pulses. In Fig.(\ref{fig1}) the initial temporal (plot (a)) and spectral (plot (b)) distributions of the pulses with temporal delay $\Delta \tau=15$ are plotted.  The initial frequencies are same for both the pulses which leads to an interference pattern. The individual spectrum of the pulses are also shown in Fig.(\ref{fig1}c) and Fig.(\ref{fig1}d) for comparison.
  
  
  \begin{figure}[h!]
  \begin{center}
  \includegraphics[trim=0.0in 2.0in 0.3in 0.3in,clip=true,  width=88mm]{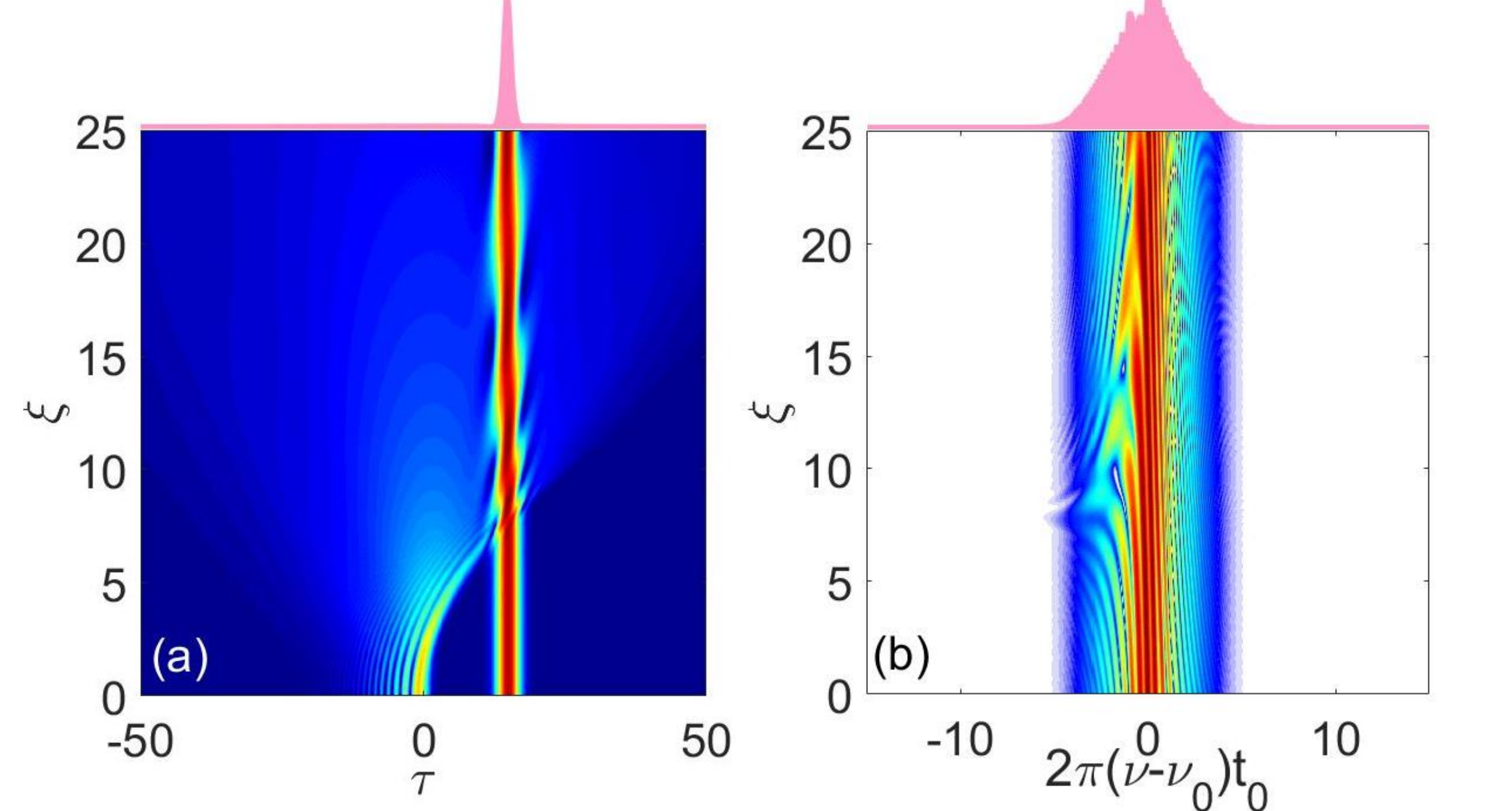}
 
  \caption{(a) Temporal  and (b) spectral  dynamics of the total field with $\delta_3=0$.  The temporal delay between the Airy pulse and soliton is $\Delta \tau=15$. The value of  $P$ is $1$  and  $a=0.1$. Note that at the collision point, $\tau_p \simeq\Delta \tau$ and there is not any radiation in the spectral evolution.}
  \label{fig2}
  \end{center}
  \end{figure}

\begin{figure}[h!]
     \begin{center}
     \includegraphics[trim=4.5in 0.0in 5.6in 0.5in,clip=true,  width=42mm]{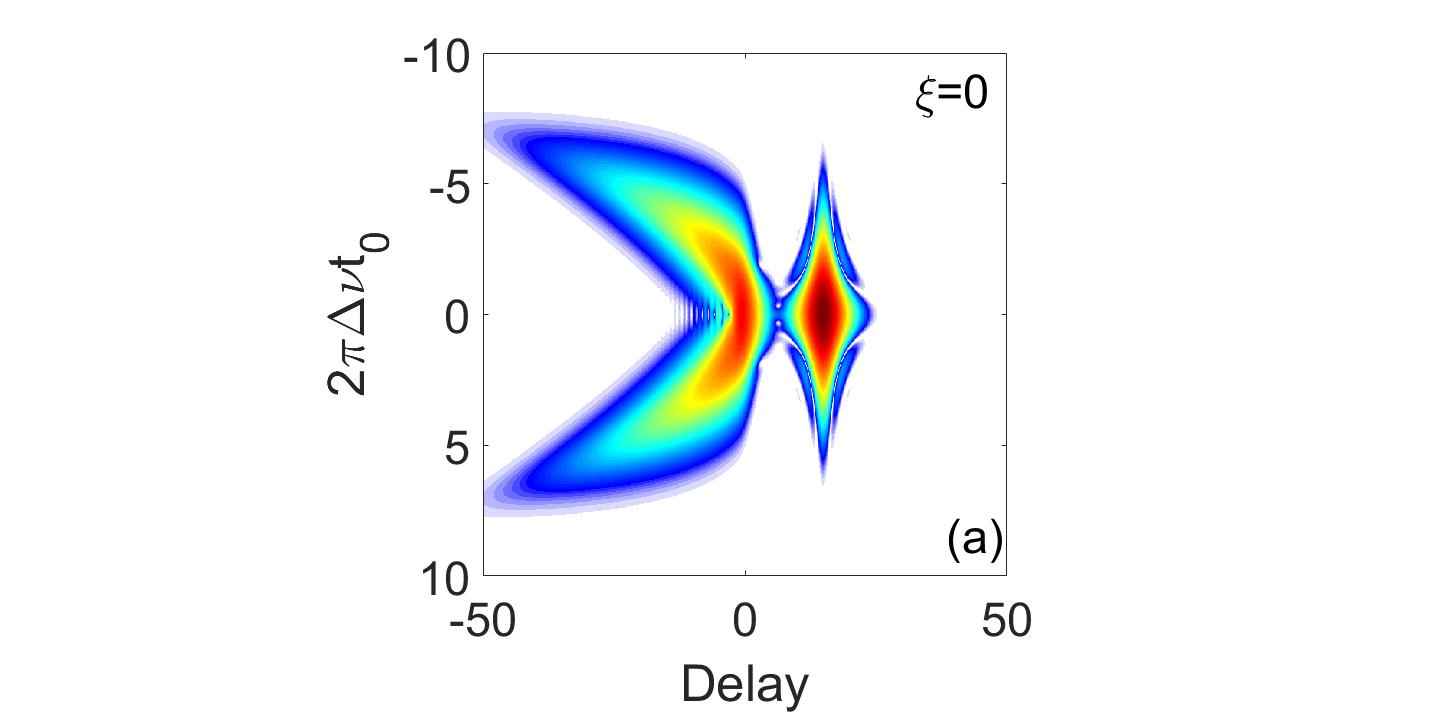}
     \includegraphics[trim=4.5in 0.0in 5.6in 0.5in,clip=true,  width=42mm]{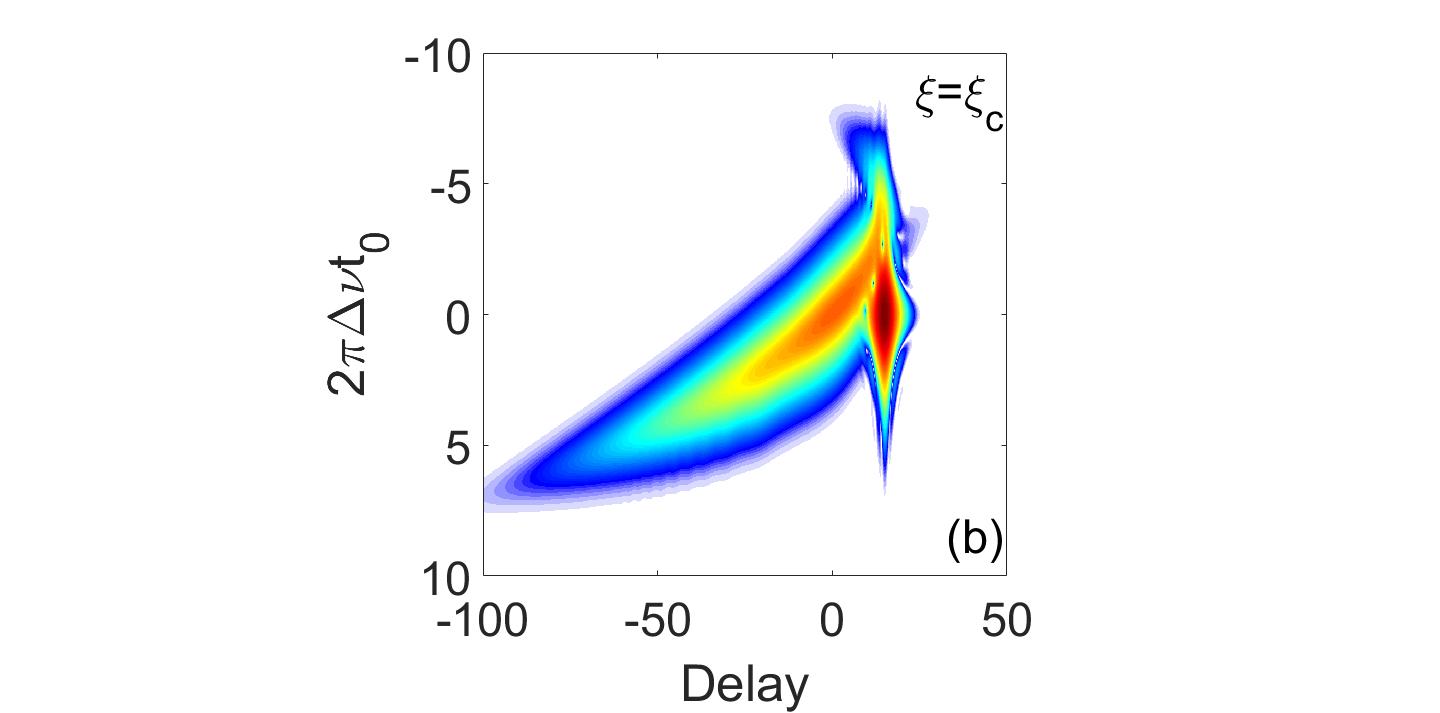}
     \includegraphics[trim=4.5in 0.0in 5.6in 0.5in,clip=true,  width=42mm]{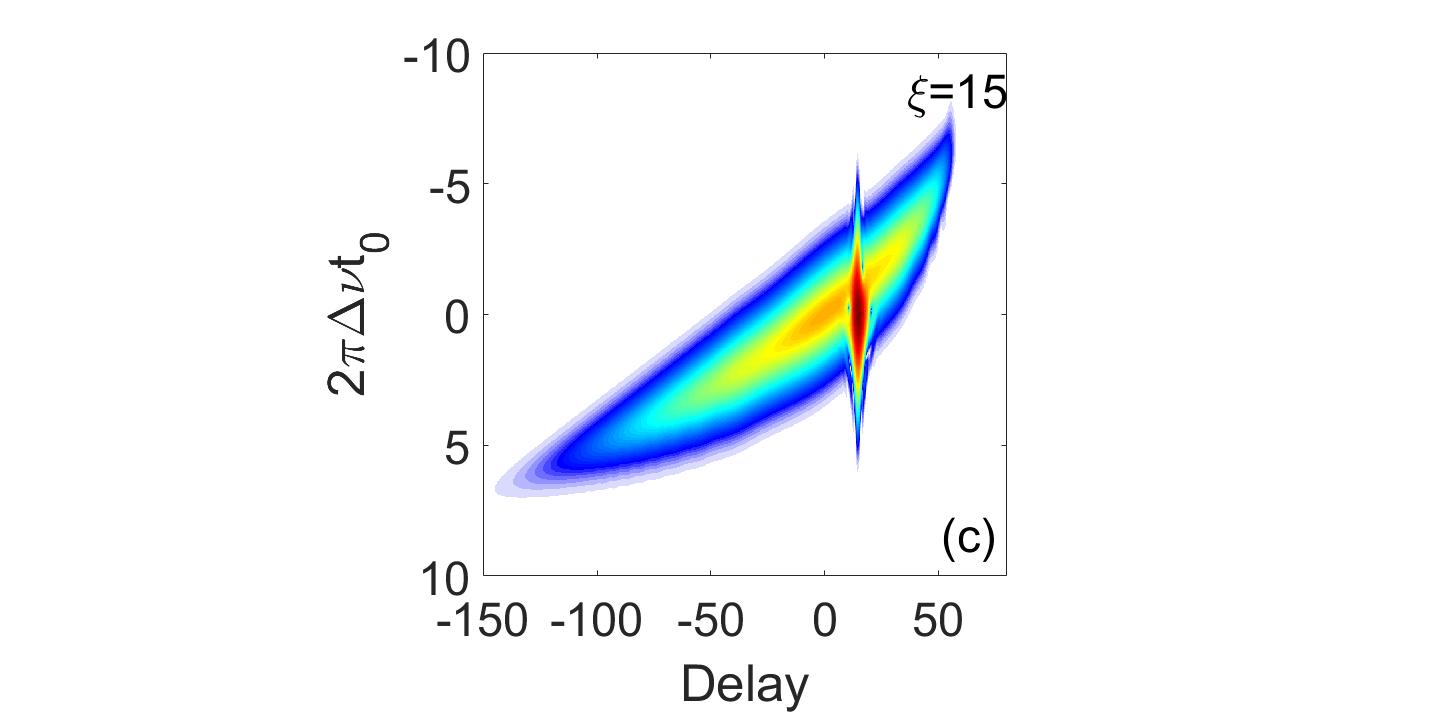}
          \includegraphics[trim=4.5in 0.0in 5.6in 0.5in,clip=true,  width=42mm]{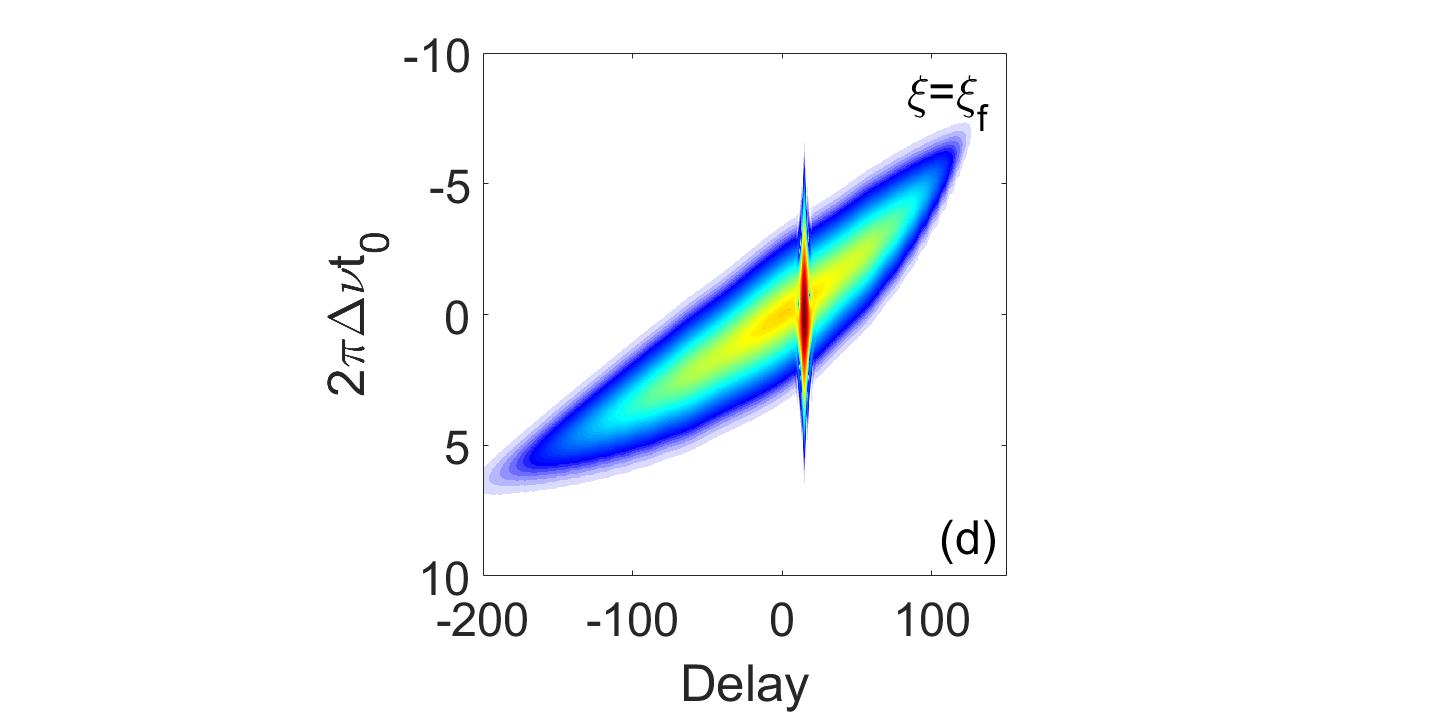}
     \caption{XFROG spectrograms at different spatial positions (a) at initial point ($\xi=0$), (b) at the collision point ($\xi=\xi_c$), (c) after the collision point ($\xi$=15) and (d) at output $(\xi$=25).}
     \label{fig3}
     \end{center}
     \end{figure}

Note, the Airy pulse is not a natural solution of the nonlinear Schr\"odinger equation and even sheds soliton at higher energy \cite{Fattal}. However, for low peak power, the Airy pulse retains it's shape even in nonlinear medium and moves with a ballistic trajectory. The main lode of the Airy pulse shifts with propagation distance as, $\tau_p\approx \tau_{0p}+\xi^2/4$, where $\tau_{0p}\approx-(3\pi/8)^{2/3}$. Hence it is easy to predict that, the decelerating Airy pulse will hit the soliton at, $\xi_c\approx 2 \sqrt{\Delta \tau-\tau_{0p}}$   . We numerically solve Eq.\eqref{q1} and show the temporal and spectral evolution in Fig.(\ref{fig2}). From these density plots it is evident that, the main lobe of the Airy pulse hits the soliton at $\xi_c \approx 8$.  This collision doesn't produce any radiation in frequency domain. To ensure that there is no collision mediated radiations, we plot the cross-correlation frequency-resolved optical grating (XFROG) diagram at four different spatial points shown in Fig.(\ref{fig3}). XFROG is a well-known technique through which we can plot the frequency
and its temporal counterpart together. Mathematically it is defined as, $s(\tau,\omega,\xi)=|\int_{-\infty}^{\infty}u(\xi,\tau')u_{ref}(\tau-\tau')e^{i\omega\tau'}|^2d\tau'$ , where $u_{ref}$ is the reference window function normally taken as the input. From the spectrograms (Fig.\ref{fig3}) we can see initially the pulses are separated in time domain where as their central frequencies are same (Fig.\ref{fig3}(a)). The pulses collide at $\xi_c\approx 8$ (Fig.\ref{fig3}(b)) without any sign of radiation. The spectrograms beyond the collision point (Fig.\ref{fig3}(c) and Fig.\ref{fig3}(d)) also do not produce any radiation patch. The Airy pulse partially penetrates the potential barrier created by the soliton. Hence, it is confirmed that the collision between the Airy pulse and the soliton  doesn't produce any radiation when only second order dispersion is present.

   
\section{Collision dynamics with third order dispersion}
   
   \noindent Now we consider the collision dynamics under higher order chromatic dispersion by introducing TOD. The governing equation of the total field in presence of TOD is given as, \cite{Agarwal_book}
    
    \begin{equation}\label{q3}
     \partial_\xi{u}=-\frac{i}{2}sgn(\beta_2)\partial^2_\tau u+\delta_3\partial^3_\tau u+i|u|^2u.          
      \end{equation}
Here $\delta_3 (=\beta_3/6|\beta_2|t_0$)  is the TOD parameter normalised by initial pulse width $t_0$, where $\beta_3$ is the TOD coefficient with unit ps$^3$/m. For anomalous dispersion $sgn(\beta_2)=-1$. It is well known that an optical soliton emits radiation when TOD perturbs the system \cite{Agarwal_book}. The radiation is either on the blue or the red side depending on the numeric sign of the TOD coefficient. The effect of TOD on the Airy pulse is rather dramatic.   Under the TOD the Airy pulse undergoes a temporal flipping and accelerates in the reverse direction \cite{driben}. Now it is interesting to study what happen when an Airy pulse collide with a soliton with non-vanishing TOD. The interaction of the soliton with the continuous wave is reported earlier and it was shown that apart from the phase matched Cherenkov radiation there are additional frequency components due to four wave mixing (FWM) between the continuous wave and the soliton\cite{Yulin}. The Airy-soliton collision may also lead to some unique radiation and our aim is to unfold the mechanism involved in this radiation process. The finite energy ($a\neq0$) Airy pulse is not a solution of the linear dispersion equation. However it is possible to derive an expression of the truncated Airy pulse moving under TOD and has the following analytic form,

   \begin{equation}\label{q4}
   u_{a}(\xi,\tau)=\frac{1}{c} \exp\left(\frac{a^3}{3}\right)Ai\left(\frac{b}{c}-\frac{n^2}{c^4}\right)\exp i\left(\frac{2n^3}{3c^6}-\frac{nb}{c^3} \right). 
   \end{equation}
   The parameters are defined as,  $c=(1-3\delta_3\xi)^{\frac{1}{3}}$;  $n=[ia+sgn(\beta_2)\xi/2]$ and $b=(\tau-a^2 )$. From the expression of the field (Eq. \refeq{q4}) it is evident that TOD significantly influences the parabolic trajectory of the Airy pulse. The numeric sign of TOD parameter ($\delta_3$) also plays an important role here, for example, a positive $\delta_3$ leads to a singularity when $c=0$. The singularity results the flipping of  the Airy pulse at a certain distance $\xi_{flip}=1/3\delta_3$. The negative $\delta_3$, on the other hand, removes the singularity (as $c\neq0$ for $\delta_3<0$ ) and helps the self-accelerating pulse to maintain its shape.

\subsection{Waveguide description}

\begin{figure}[h!]
           \begin{center}
           \includegraphics[trim=0.5in 0.0in 0.25in 0.05in,clip=true,  width=80mm]{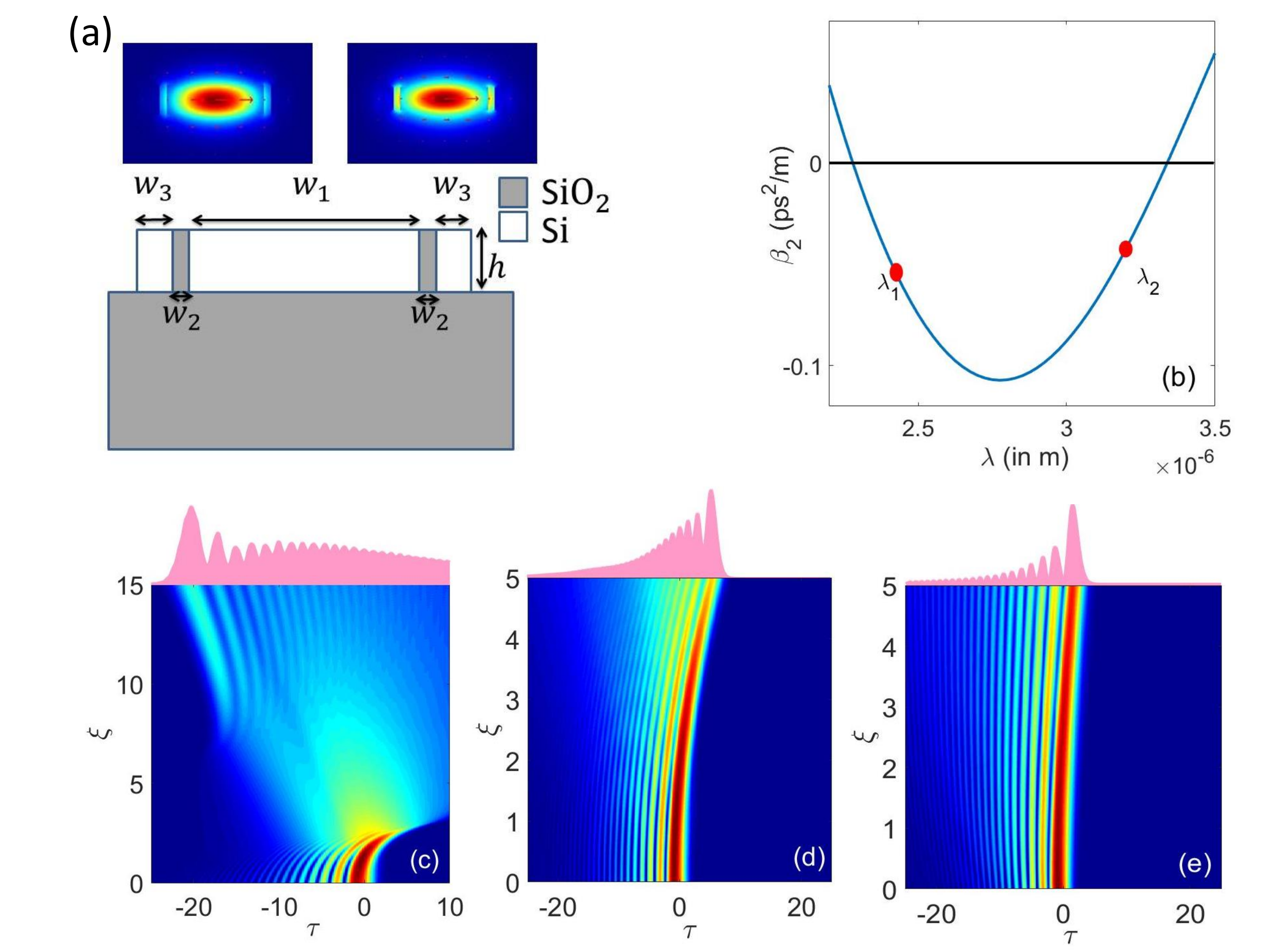}
           \vspace{1em}
            \caption{a) The block diagram and geometry of the proposed Si-based waveguide. The dimensions are $w_1$ = 1500 nm; $w_2$ = 110 nm; $w_3$ = 230 nm  and $h$ = 410 nm. The confinement of the fundamental TE mode at the operating wavelengths ($\lambda_1=$ 2425 nm and $\lambda_2=$3200 nm) are shown in the inset. (b) Dispersion profile of the waveguide. The operating wavelengths are indicated by the red solid dots. For $\lambda_1$ the TOD parameter is positive and for $\lambda_2$  TOD parameter is negative. Dynamics of a truncated Airy pulse for (c)  $\delta_3>0$(=0.08) (d)  without TOD ($\delta_3=0$) and (e) negative $\delta_3<0$(=-0.08).   }
           
           \label{fig4}
           \end{center}
           \end{figure}

 \noindent To realize the effect of the positive and negative TOD on the Airy pulse dynamics we propose a realistic Si-based waveguide that exhibits two zero dispersion wavelengths. The waveguide geometry and related dispersion profile is shown in Fig.(\ref{fig4}a) and Fig.(\ref{fig4}b) respectively. Note that, for two zero dispersion profile the slope of GVD changes its numeric sign across the lowest dispersion value. One can achieve the desired $\delta_3$ values with opposite signs at two different launching wavelengths as indicated by the red dots in Fig.(\ref{fig4}b).

 \noindent For the proposed waveguide, we have evaluated the values of GVD parameter ($\beta_2$) and TOD parameter ($\beta_3$) in real unit by using COMSOL multi-physics software. The values of GVD and TOD coefficients are $\beta_2$=-0.038  ps$^2$/m and $\beta_3$=0.0011 ps$^3$/m respectively at the first launching wavelength $\lambda_1$= 2425 nm. For the second launching wavelength, $\lambda_2$=3200 nm, the values are $\beta_2$=-0.043  ps$^2$/m and $\beta_3$=-0.0015 ps$^3$/m. In our simulation we use the value of $\delta_3$ in the range 0.05 to 0.1 which we can achieve for the width of the pulse between 50 fs to 100 fs. (e.g. for $\lambda_1$, $\delta_3 \approx$ 0.08  for $t_0 =$ 60 fs ; similarly, for $\lambda_2$, $\delta_3 \approx$ -0.08  for $t_0 =$ 75 fs). The nonlinear parameter ($\gamma_r$) is defined as $\gamma_r=k_0n_2/A_{eff}$, where $k_0=2\pi/\lambda_0$, $n_2$ is the Kerr coefficient and $A_{eff}$ is the effective area. For Si, $n_2\approx3\times10^{-18} m^2W^{-1}$. The effective area of the confined mode ($A_{eff}=(\iint\limits_{-\infty}^{+\infty}|\psi(x,y)|^2dxdy)^2/\iint \limits_{-\infty}^{+\infty}|\psi(x,y)|^4dxdy$) is calculated as, 0.51 $\mu m^2$ for $\lambda_1$ and 0.66 $\mu m^2$ for $\lambda_2$. To have an idea about the nonlinear effect on the dynamics of the pulses we have compared the values of $L_D (=t_0^2/|\beta_2|)$ and $L_{NL} (=1/\gamma_rP_0)$. Note that we have to be careful enough in choosing the external parameters like peak power, pulse width etc, to excite soliton and Airy pulse together inside the waveguide. In Kerr medium the Airy pulse sheds optical soliton for high input power and loses its characteristic shape \cite{Fattal}. To retain the shape of the Airy pulse in Kerr medium, we need to manipulate the input power judiciously. The relative nonlinear effect is reduced or becomes negligible when $L_{NL}>>L_{D}$. It can be easily seen that, for initial pulse width $t_0=$ 100 fs and peak power $P_0$ $\approx$ 200 mW the value of $L_D \approx$ 0.26 m  and $L_{NL} \approx$ 0.33 m. On the contrary, for peak power $P_0$ $\approx$ 40 mW, $L_D \approx$ 0.26 m  and $L_{NL} \approx$ 1.65 m which satisfies the condition $L_{NL}>>L_{D}$. The dynamics of a truncated Airy pulse in Kerr medium with different dispersion parameters are shown in Fig. \ref{fig4}(c)-(e). In Fig.\ref{fig4}(c) the temporal dynamics is shown for $\delta_3>0$(=0.08).It is evident that, the dynamics is not uniform and the pulse experiences a temporal flipping with reverse acceleration. The flipping distance $\xi_{flip}(\propto 1/\delta_3)$, reduces significantly when TOD is very high. In Fig. \ref{fig4}(d) and \ref{fig4}(e) the dynamics is shown for $\delta_3$=0 and $\delta_3 <0$(=-0.08). It is evident from the final field distributions (attached within the figures)  that the shape of the Airy pulse is more robust for negative $\delta_3$ (Fig.\ref{fig4}(e)) than the pulse which does not face any higher order dispersion (Fig.\ref{fig4}(d)) .

   \subsection{Collision dynamics under negative TOD ($\delta_3<0$)}

\begin{figure}[h!]
  \begin{center}
  \includegraphics[trim=0.0in 1.9in 0.1in 0.2in,clip=true,  width=86mm]{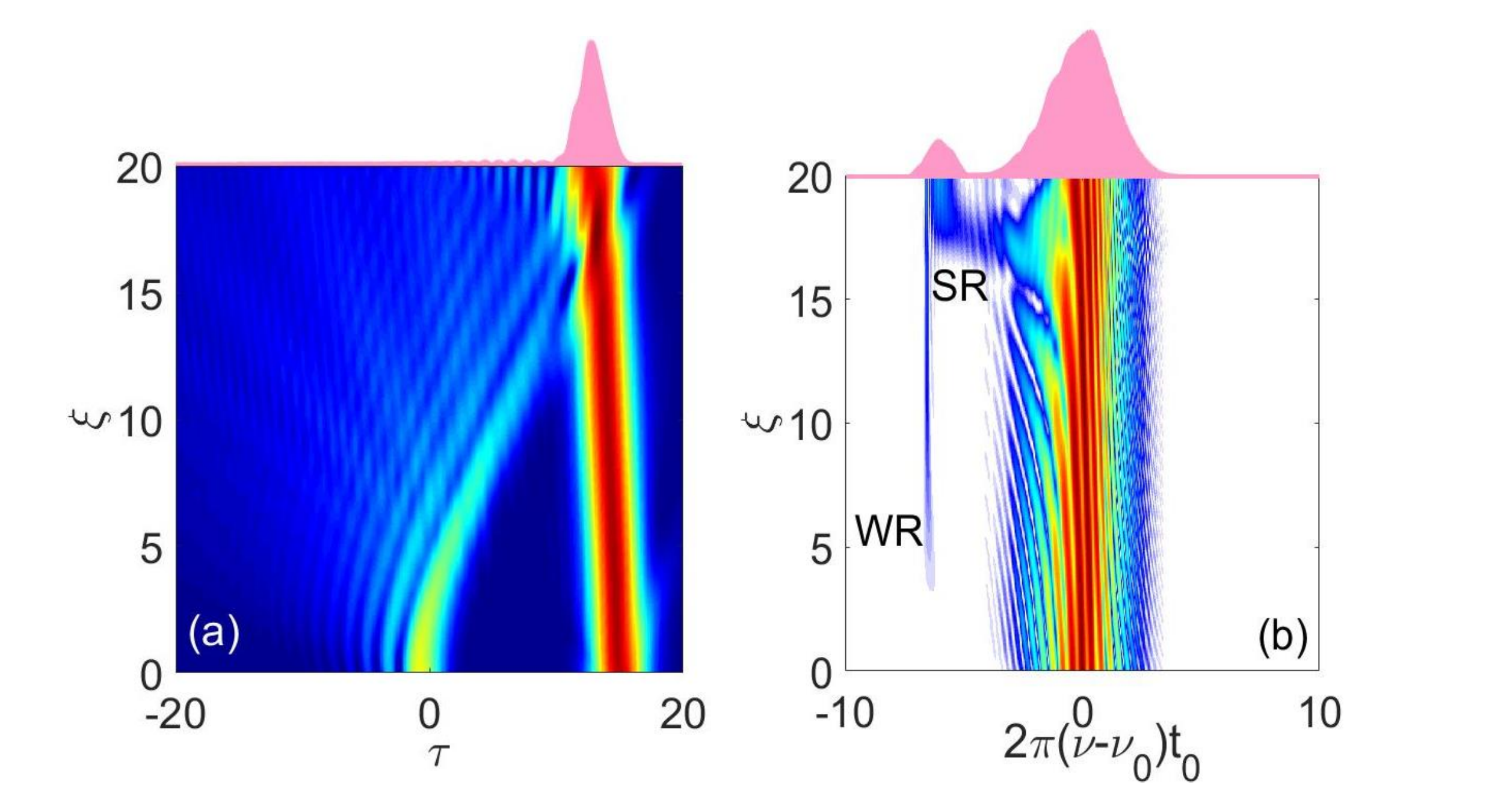}
\caption{ The (a)temporal  and (b)spectral  dynamics of the pulses with negative $\delta_3$ (= -0.08). The initial separation between the pulse  $\Delta \tau=15$. For airy pulse  $a=0.25$.}  
  \label{fig5}
  \end{center}
  \end{figure}

\noindent It has already been shown that higher order dispersion influences the ballistic trajectory of an Airy pulse. From the general solution Eq.\eqref{q4},  it can be shown that for negative $\delta_3$ the peak of the main lobe of Airy pulse moves with a delay rate $\frac{d\tau}{d\xi}\approx \frac{2\xi+3|\delta_3|\xi^2}{4c^6}$, where $c=(1+3|\delta_3|\xi)^{1/3}$. We launch the Airy pulse and soliton together where the soliton is delayed by  $\Delta\tau=15$ ensuring no temporal overlap. The presence of negative TOD preserves the unique nature of the Airy pulse. The peak power of the main lobe of the Airy pulse is such that, it does not experience any nonlinear effects.  In Fig.(\ref{fig5}) we  plot the temporal and spectral dynamics of the total field. The collision between the soliton and the Airy pulse leads to a strong radiation (SR) in the frequency domain shown in Fig.(\ref{fig5}b). If we look carefully, apart from the strong radiation, a weak radiation (WR) is observed in the spectral evolution. Even though the spectral locations of the strong and weak radiations are close to each other, they are not same as the weak radiation appears much earlier than the strong radiation (see Fig.(\ref{fig5}b) ).  This weak radiation can be linked with well known \textit{Chereknov radiation} which appears when an optical soliton is perturbed by TOD \cite{Agarwal_book} . Note, the situation shown in Fig.(\ref{fig2}) and Fig.(\ref{fig5}) is identical except the fact that in the later case we include TOD which results radiations. It is interesting to observe that in Airy-soliton system the \textit{strong radiation} appears only when the pulses collide with each other under TOD. The radiation ceases to exist either there is no temporal collision or $\delta_3$ is zero.

\begin{figure}[h!]
  \begin{center}
  \includegraphics[trim=4.5in 0.05in 5.6in 0.5in,clip=true,  width=42mm]{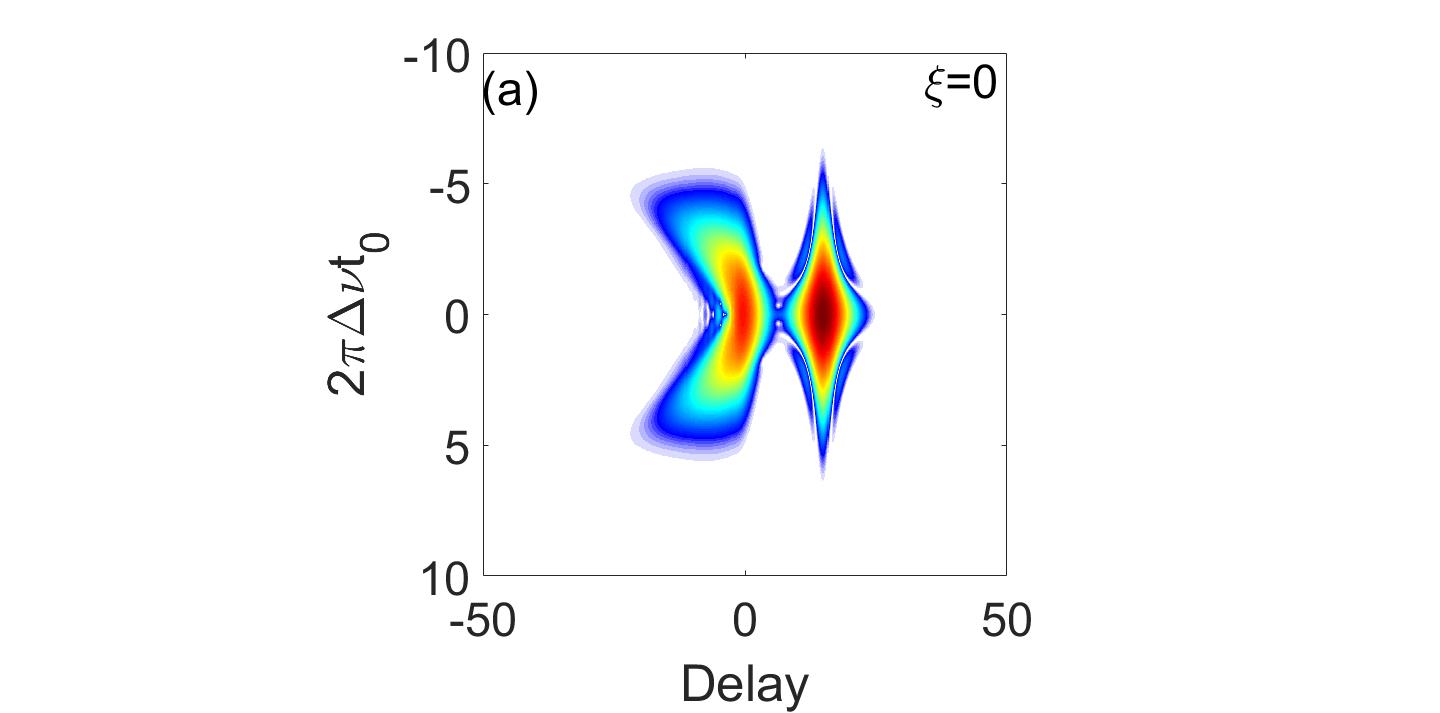}
  \includegraphics[trim=4.5in 0.05in 5.6in 0.5in,clip=true,  width=42mm]{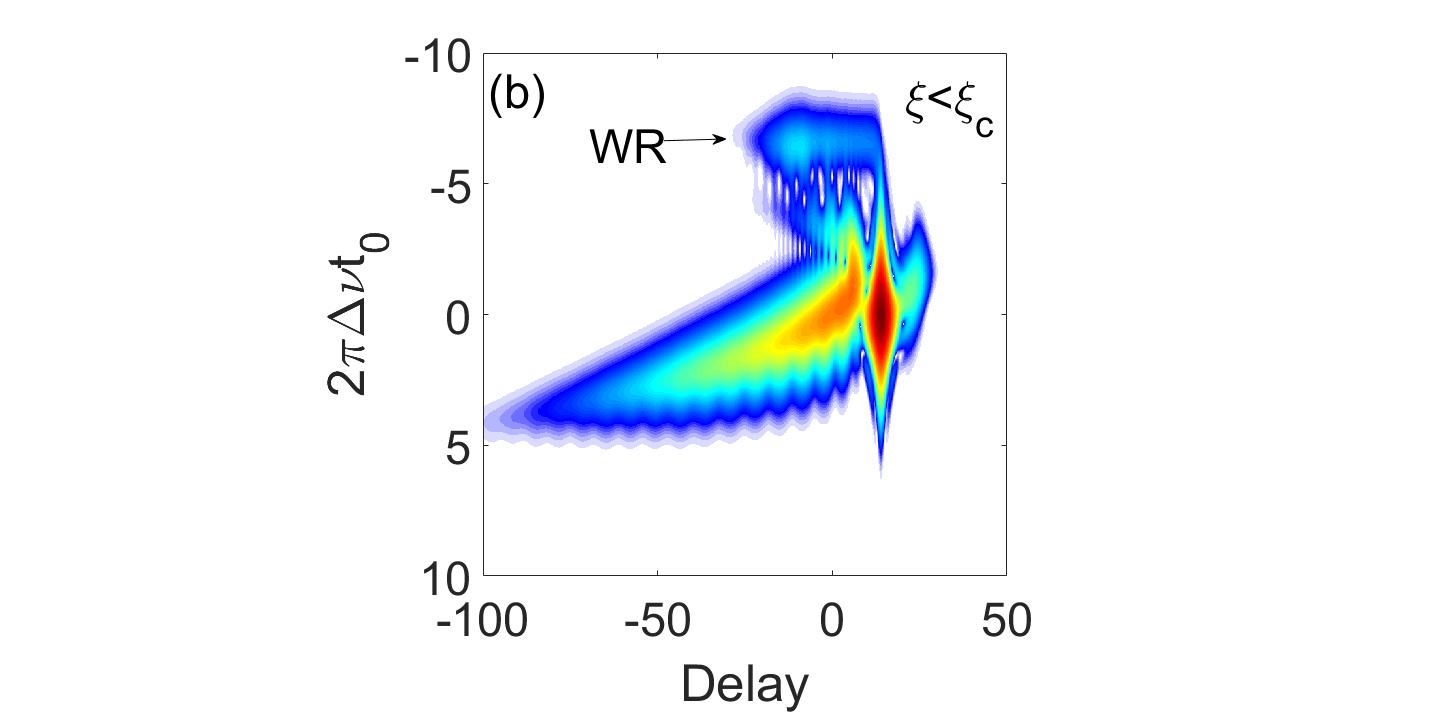}
  \includegraphics[trim=4.5in 0.05in 5.6in 0.5in,clip=true,  width=42mm]{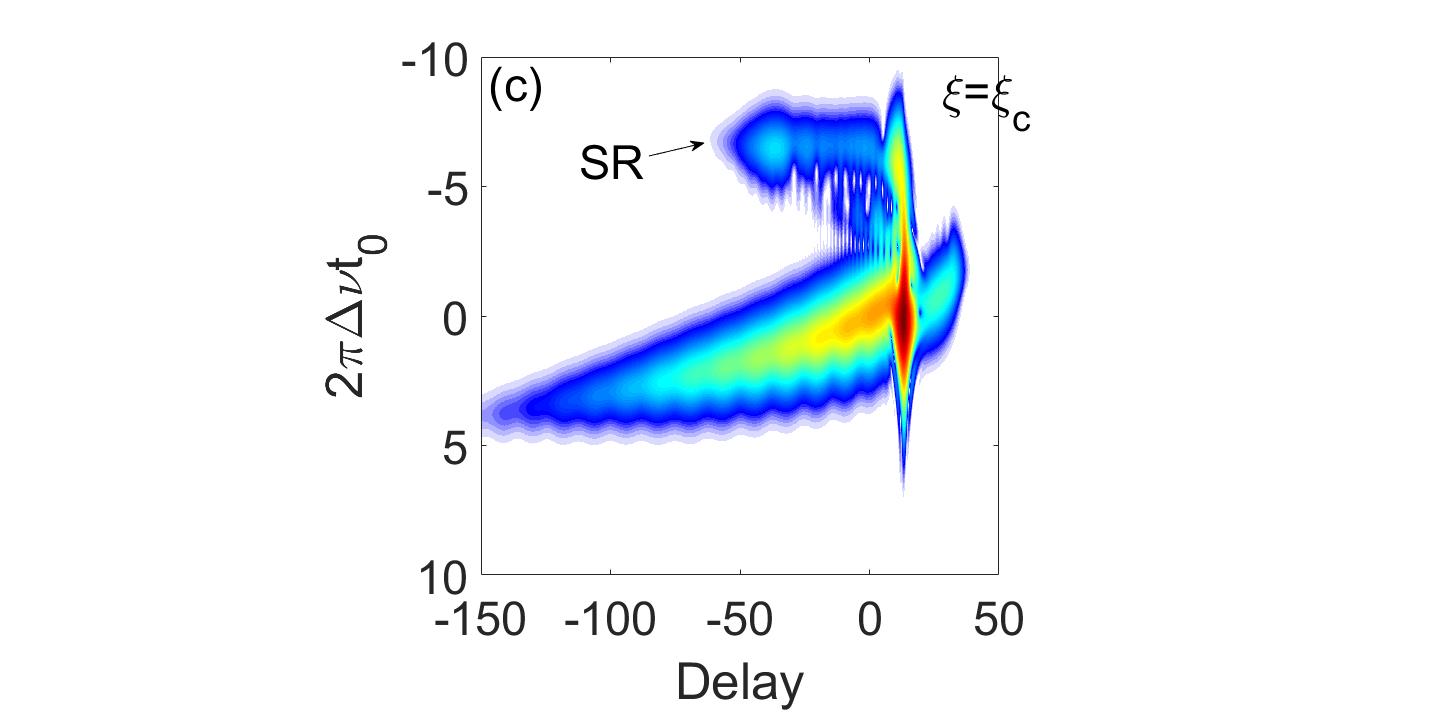}
  \includegraphics[trim=4.5in 0.05in 5.6in 0.5in,clip=true,  width=42mm]{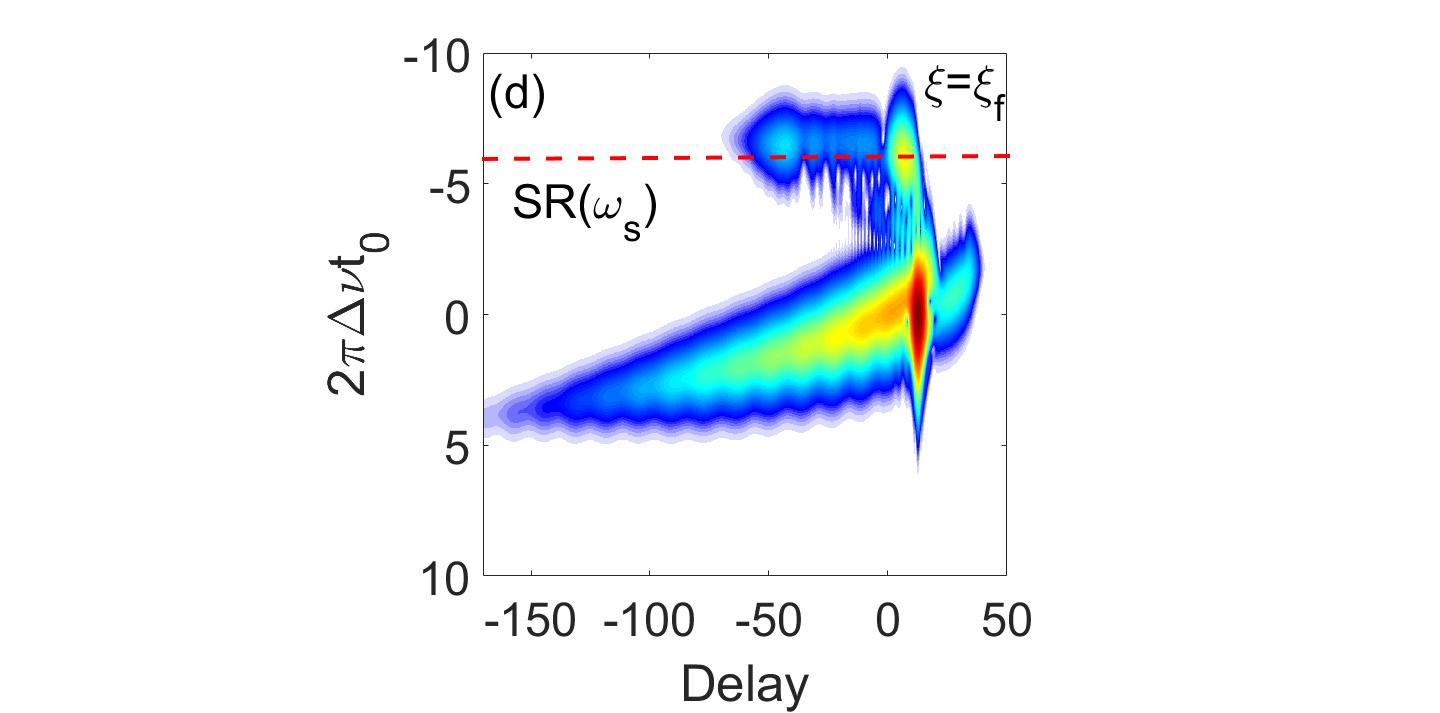}
  
  \caption{ The XFROG spectrograms at different distances.  (a) At $\xi=0$ where two pulse are temporally separated ($\Delta \tau=15$). (b) At $\xi=10$ ($\xi<\xi_c$) where only the soliton mediated  weak radiation is visible. (c) At the collision point ($\xi=\xi_c$) where Airy pulse collide with soliton and leads to a strong radiation. (d) At output ($\xi=\xi_f$) where the coexistence of strong (indicated by the red dotted line) and weak  radiation is evident.  The value $\delta_3=-0.08$, and the truncation parameter $a=0.25$.  }
  \label{fig6}
  \end{center}
  \end{figure}

\begin{figure}[h!]
  \begin{center}
  \includegraphics[trim=1.5in 0.0in 1.9in 0.0in,clip=true,  width=92mm]{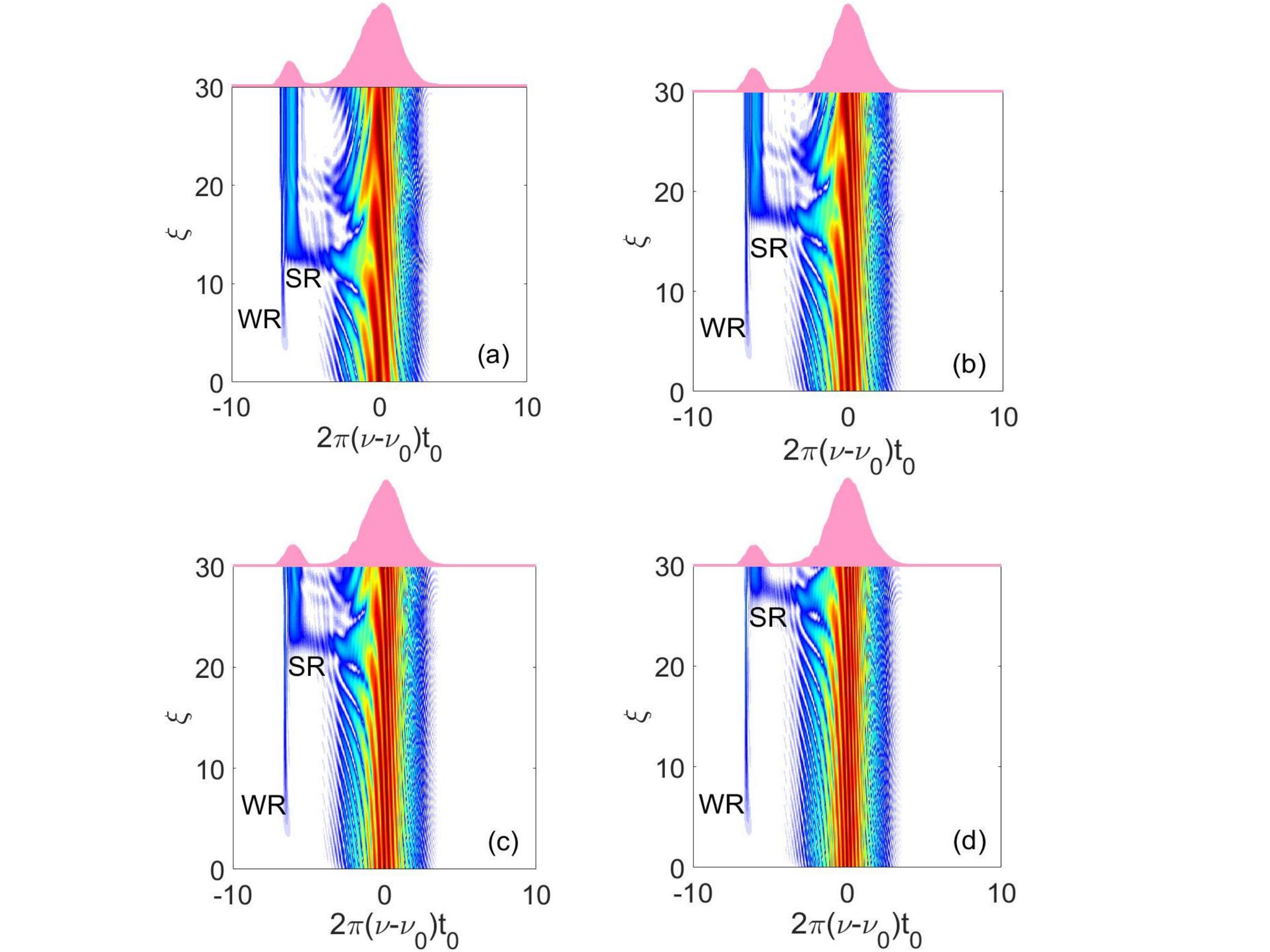}
\caption {The spectral dynamics of the pulses for a fixed TOD parameter ( $\delta_3=-0.08$) with different temporal separations (a) $\Delta \tau=10$, (b)$\Delta \tau=15$, (c)$\Delta \tau=20$ and (d)$\Delta \tau=25$. The radiation is generated at different collision points based on the initial separation of the pulses. However the spectral location of the radiation is hardly affected by delay $\Delta \tau$ . }
  \label{fig7}
  \end{center}
  \end{figure}

The XFROG spectrogram plots (Fig \ref{fig6}) highlight the dynamics with more clarity. In Fig.{\ref{fig6}(a-d)} we show four  different stages of the co-propagating Airy-soliton dynamics. If we define the collision distance as $\xi_c$ then, in plot (a) and (b) we capture the scenario before collision ($\xi<\xi_c$). A faint patch in plot (b) indicates the weak radiation generated by the soliton under TOD. The spectrogram at the collision point (plot(c)) confirms the presence of a new spectral component which is much more stronger than the earlier radiation. Finally in plot (d) we show the situation beyond collision point ($\xi>\xi_c$). It is observed that, the energy of the Airy pulse is mostly reflected back from the barrier created by the strong soliton pulse. The red dotted line in Fig. \ref{fig6}(d) indicates the spectral position ($\omega_s$) of the collision mediated radiation.  Since the collision point $\xi_c(\Delta \tau)$, is a function of the delay, we can control the radiation by manipulating $\Delta \tau$, keeping $\delta_3$ fixed. In Fig.\ref{fig7}(a-d) we plot the spectral evolution of the co-propagating pulses with different delay time $\Delta \tau$. The  plots show that, the spatial location of the strong radiation shifts significantly with $\Delta \tau$, where as the weak \textit{Cherenkov} radiation appears at a fixed spatial position invariably. It is expected because WR is solely controlled by TOD and not depends on delay $\Delta \tau$.  We also observe that the spectral location of the strong radiation hardly shifts with $\Delta \tau$. From all the above results it is evident that  the interaction between the co-propagating Airy pulse and the soliton in time domain leads to a strong radiation. It is also confirmed that, this collision mediated radiation vanishes if $\delta_3=0$. In the following section we try to propose a theoretical analysis that can explain the origin of this strong radiation.
\\

\noindent \textbf{ Theory of collision-assisted radiation for $\delta_3<0$} \\

Under TOD the soliton emits the radiation at a frequency for which the propagation constant of the linear wave matches the input pulse momentum \cite{Dmitry}. Such radiation is called the \textit{Chrenkove radiation}. In co-propagating Airy-soliton system, the phase matching condition is modified due to the collision between the self-accelerating Airy pulse and the soliton. This collision  leads to a strong radiation.  The total field of the system  can be written as, $u=u_{s}+g$, where $u_{s}=\bar{u}_{s}\exp(i\phi_{s})$ is the soliton field and $g$ is the superposition of all the other waves including the Airy pulse. Putting $u$ into Eq.(\ref{q3}) and linearising $g$ we get,

\begin{equation}\label{q5}
       \partial_{\xi}{g}-\hat{D}(i\partial_{\tau}) g=\delta_3\partial^3_\tau u_s+i2\bar{u}_{s}^2 g+i\bar{u}_{s}^2 g^*e^{2i\phi_{s}},          
       \end{equation}
where the dispersion operator $\hat{D}(i\partial_{\tau})=\frac{i}{2}\partial_\tau^2+\delta_3\partial_\tau^3$.  The term $g$ comprises of the the airy pulse and linear dispersive wave  as, $g=u_{a}+\psi$, where $u_{a}=\bar{u}_a \exp(i\phi_a)$ and $\psi=\bar{\psi}\exp(i\phi_c)$. Using the explicit form of $g$ in Eq.(\ref{q5}) we may have the governing equation of the radiation wave as,

\begin{equation}\label{q6}
       \partial_\xi\psi-\hat{D}(i\partial_{\tau})\psi= \delta_3\partial^3_\tau u_{s}+i\bar{u}_{s}^2\bar{u}_ae^{i(2\phi_{s}-\phi_a)} +i2\bar{u}_{s}^2\bar{u}_ae^{i\phi_a}      
       \end{equation}
To obtain the radiation modes we neglect the higher order terms of $u_s$ associated with the field $\psi$. The wave number matching with the driving terms in the right-hand side leads to the phase matching condition of the strong radiation as,  
\begin{equation}\label{q7}
\phi_c=2\phi_{s}-\phi_a
\end{equation}
 For small truncation parameter we can approximate the Airy pulse solution (shown in Eq.(\ref{q4})) as, 

\begin{equation}\label{q8}
   u_{a}(\xi,\tau)\approx \frac{1}{c}Ai \left(\frac{\tau}{c}-\frac{\xi^2}{4c^4}\right)e^{i \phi_{a}}, 
   \end{equation}
where the phase has the following form,

\begin{equation}\label{q9}
\phi_{a}=\frac{\Gamma}{c^6},
\end{equation}
with $\Gamma (\xi,\tau)=[\frac{\xi}{2}\tau-\frac{3}{2}\delta_3\tau\xi^2-\frac{\xi^3}{12}]$. The temporal position of the main lobe of the Airy pulse is given as,
\begin{equation}\label{q10}
\tau_{p}\approx\frac{\xi^2}{4(1-3\delta_3\xi)}.
\end{equation}
From the above solution  we can see that the pulse experiences a singularity for a positive TOD parameter at, $\xi_{flip}=(3\delta_3)^{-1}$. However this singularity can be avoided when the pulse is propagating under negative TOD ($\delta_3<0$). If we neglect the small  temporal shift of the soliton under TOD, we can develop a relationship between the collision point $\xi_{c}$  and the initial delay $\Delta \tau$ as,  
\begin{equation}\label{q11}
\xi_c=\xi_0 \left[-sgn(\delta_3)+\sqrt{(1+\mu^2)}\right];
\end{equation}
where $\xi_0=6{|\delta_3|}\Delta \tau$ and $\mu=(3\delta_3\sqrt{\Delta \tau})^{-1}$. Note that, for the limit $\delta_3\rightarrow0$, $\xi_c=2\sqrt{\Delta \tau}$ and at the collision point, $\tau_p\simeq \Delta \tau$. Exploiting Eq.\eqref{q9}-\eqref{q11}, it can be shown that, at the collision point, $\phi_{a}=\xi_c^3/24c_c^6$, where $c_c=(1-3\delta_3\xi_c)^{1/3}$. Simplifying the phase-matching condition shown in Eq.\eqref{q7}, we can derive the detuned frequency of collision-mediated strong radiation ($\omega_s$) as,
\begin{equation}\label{q12}
\omega_{s}=\frac{1}{2\delta_3}+4\delta_3-4\delta_3\frac{\phi_{a}}{\xi}\bigg\vert_{\xi=\xi_c}
\end{equation}
A careful investigation of Eq.\eqref{q12} reviles interesting facts. Note that, the third term of the right-hand side of the Eq.\eqref{q12} is the signature of the collision mediated radiation and it vanishes when there is no collision ($\xi_c\rightarrow \infty$). The third term can also vanish when $\delta_3=0$ , i.e in spite of the collision, without TOD $(\delta_3=0)$, there will be no radiation , which is consistent with our earlier findings. The detuned frequency of the weak radiation ($\omega_w$) can be obtained using the relation $D(\omega_w)=1/2$, where $D(\omega_w)=-\omega_w^2/2+\delta_3\omega_w^3$  \cite{Skryabin}.

 \begin{figure}[h!]
    \begin{center}
    \includegraphics[trim=6.0in 0.0in 7.1in 0.0in,clip=true,  width=42mm]{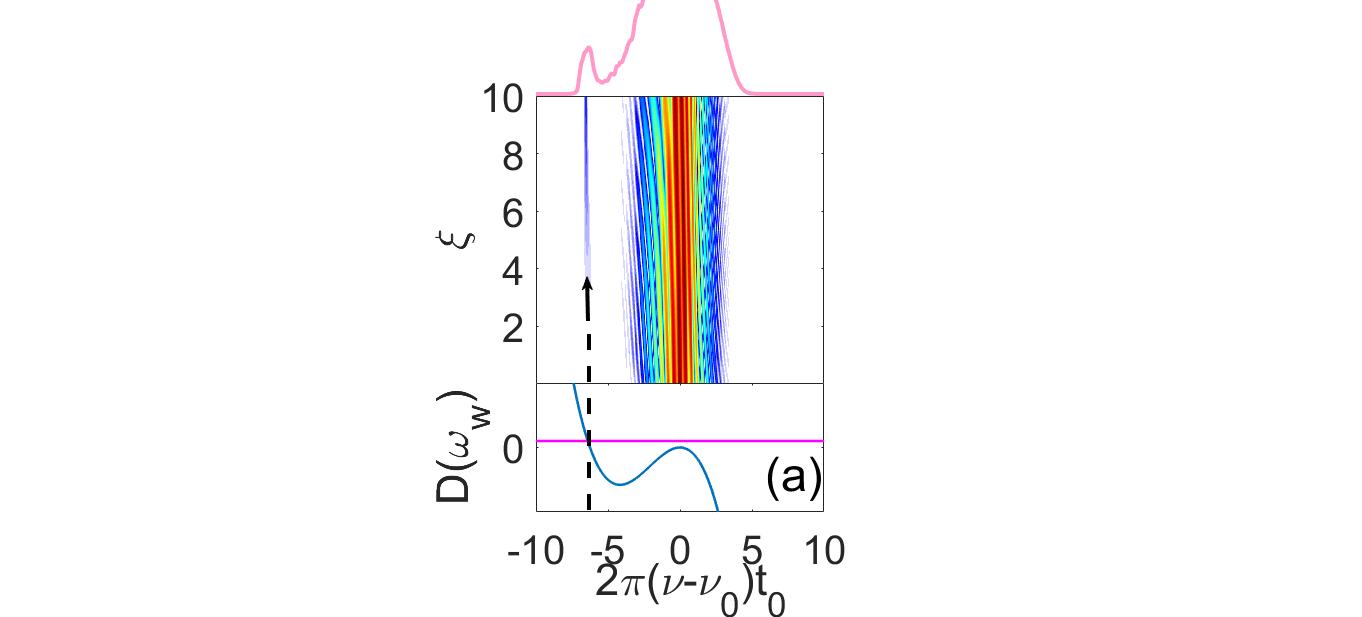}
    \includegraphics[trim=6.0in 0.0in 6.7in 0.0in,clip=true,  width=43mm]{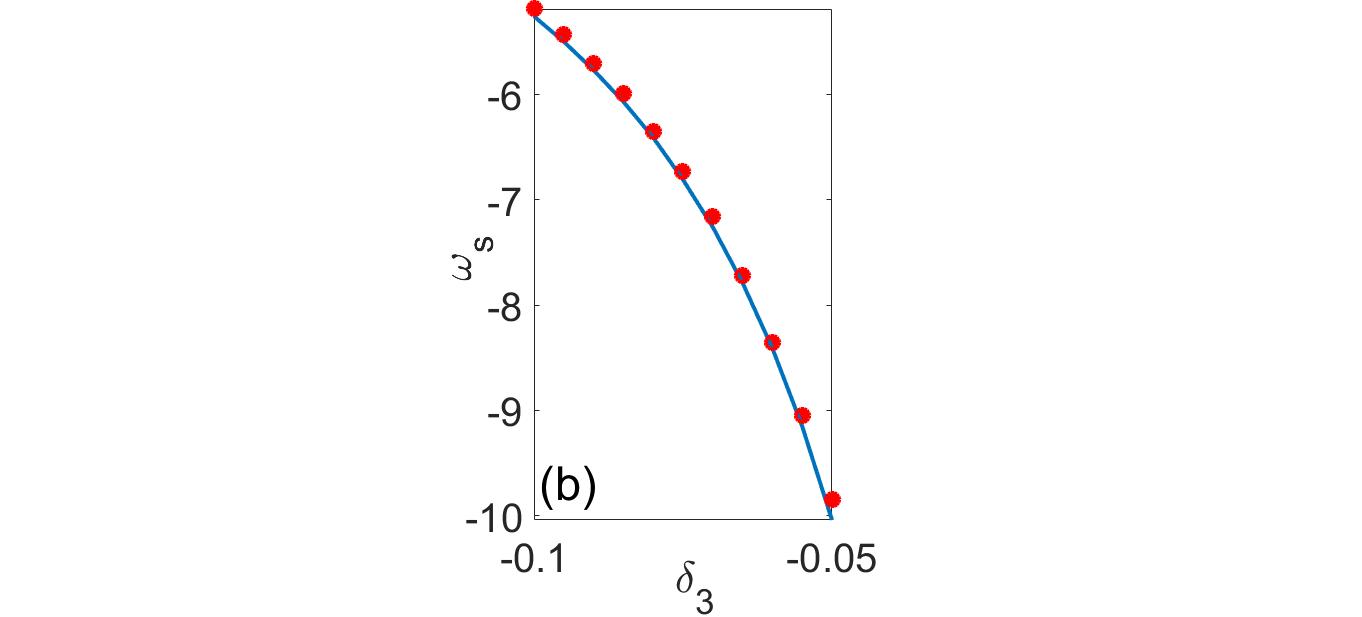}
   
    \caption {(a) The spectral dynamics  before the collision point($\xi<\xi_c$) with $\delta_3=-0.08$. The spectral position of the weak radiation is verified by plotting the phase matching curve for the weak radiation.The final spectral distribution is inserted at the top of the figure. (b) The comparison of analytically and numerically found results for all the frequencies obtained due to the perturbations present in the system. }
    \label{fig8}
    \end{center}
 \end{figure}
 
\begin{figure}[h!]
      \begin{center}
      \includegraphics[trim=0in 1.8in 0.00in 1.1in,clip=true,  width=90mm]{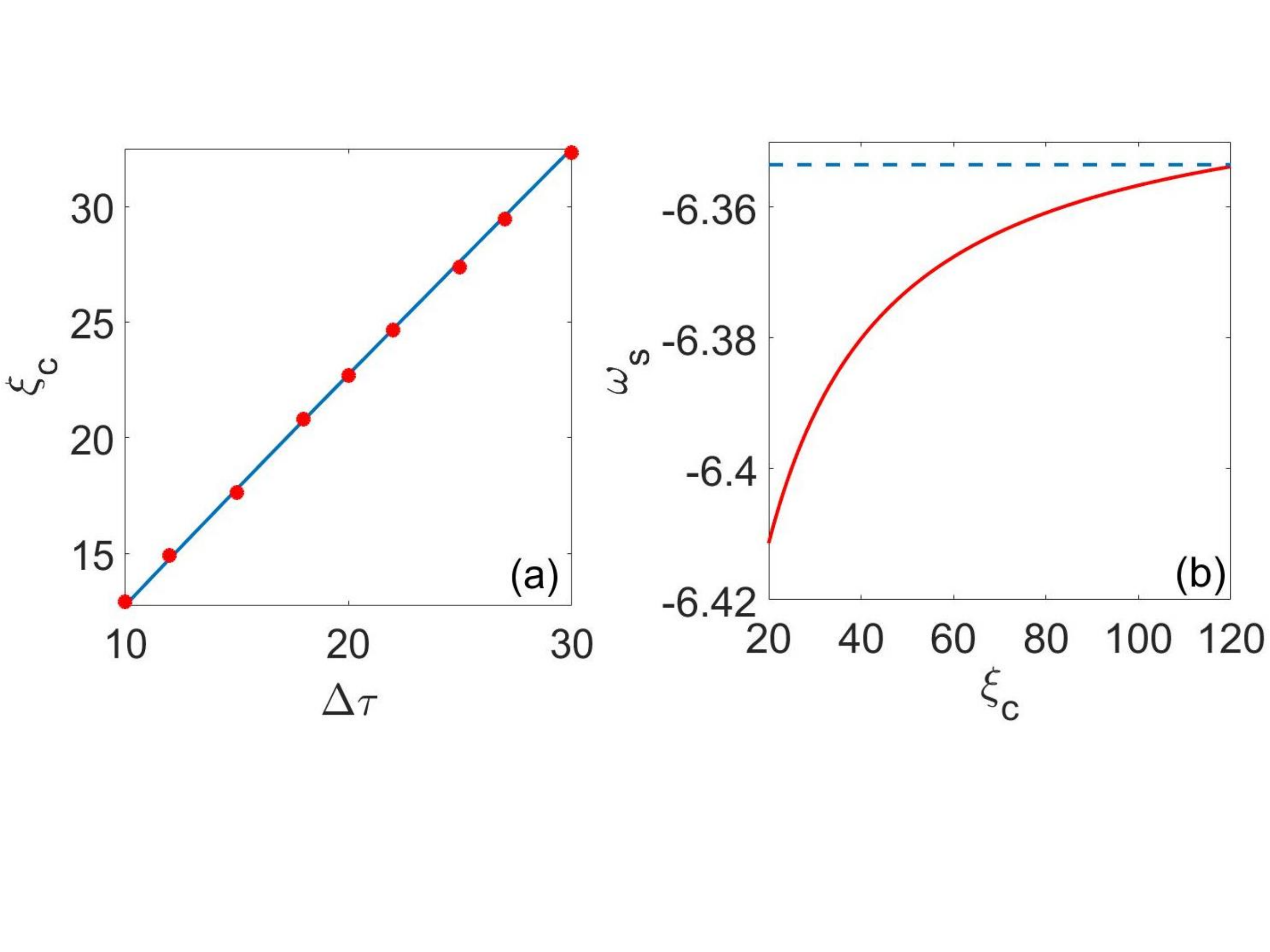}
     
      \caption {(a) Collision distance ($\xi_c$) as a function of  initial temporal separation ($\Delta \tau$) between the pulses. The analytical formula (blue solid line) is verified by numerically data (red dots).  (b) The variation of collision-mediated strong radiation frequency $(\omega_s)$ with $\xi_c$ (for fixed $\delta_3=-0.08$). The radiation frequency $\omega_s$ hardly changes (around 1$\%$) with $\xi_c$.}
      \label{fig9}
      \end{center}
      \end{figure}

 In  Fig.\ref{fig8}(a) we have shown the evolution of the weak radiation whose spectral location can be predicted efficiently by the phase matching condition which leads to the radiation frequency as, $\omega_w=2\delta_3+(2\delta_3)^{-1}$ . In Fig.  \ref{fig8}(b) we have plotted the spectral location of the collision-mediated strong radiation as a function of $\delta_3$. The solid line represents the expression proposed by us in Eq.(\ref{q12}) which corroborate well with the numerical data shown by solid red coloured dots. There is another important  fact that can be noticed from Eq.(\ref{q12}). It  seems the collision mediated radiation frequency ($\omega_s$) depends on the  collision point ($\xi_c$) and in that case one can simply tailor the radiation by changing the soliton delay time $\Delta \tau$   using Eq.(\ref{q11}). In Fig.\ref{fig9}(a) we have shown the linear relationship between collision point ($\xi_c$) and temporal delay ($\Delta \tau$) between pulses. This contradicts the numerically found results where we have shown that the spectral position of the collision mediated radiation hardly depend on the position of the collision between the pulses. In Fig.\ref{fig9}(b) we have plotted the variation of the analytically found expression of $\omega_s$ Eq.(\ref{q12}) as a function of the collision distance $\xi_c$. From the figure it is clear that the value of $\omega_s$ does not vary much over the distance and it saturates to a fixed value after a certain distance which is indicated by the blue dashed line in Fig.\ref{fig9}(b). This value is very close to the value of the spectral position of the weak radiation. 


 \subsection{Collision dynamics under positive TOD ($\delta_3>0$)}
\noindent It is already shown that, the strong radiation appears when the self accelerating Airy pulse interacts with a delayed soliton in the environment of negative TOD. In absence of TOD, the collision doesn't lead to any radiation. The Airy pulse experiences a singularity at $\xi= (3\delta_3)^{-1}$ when TOD is positive ($\delta_3>0$). In such case, the Airy pulse undergoes a temporal flipping and accelerates in the reverse direction. The location of the temporal flipping is generally defined as, $\xi_{flip}$ which is inversely related to the strength of the TOD parameter. Now for $\delta_3>0$ the Airy pulse interacts with a delayed soliton in two possible ways,  (i) before the flipping point and (ii) after the flipping point.  Note, in order to achieve the collision after the flipping point, the soliton needs to be launched in advance since Airy pulse accelerates in reverse direction beyond the flipping point. In our study we separately investigate the dynamics for both situations.

   \subsubsection{Dynamics before the flipping point ($\xi<\xi_{flip}$)}
   
\begin{figure}[h!]
    \begin{center}
 \includegraphics[trim=0.5in 0.01in 1.5in 0.05in,clip=true,  width=100mm]{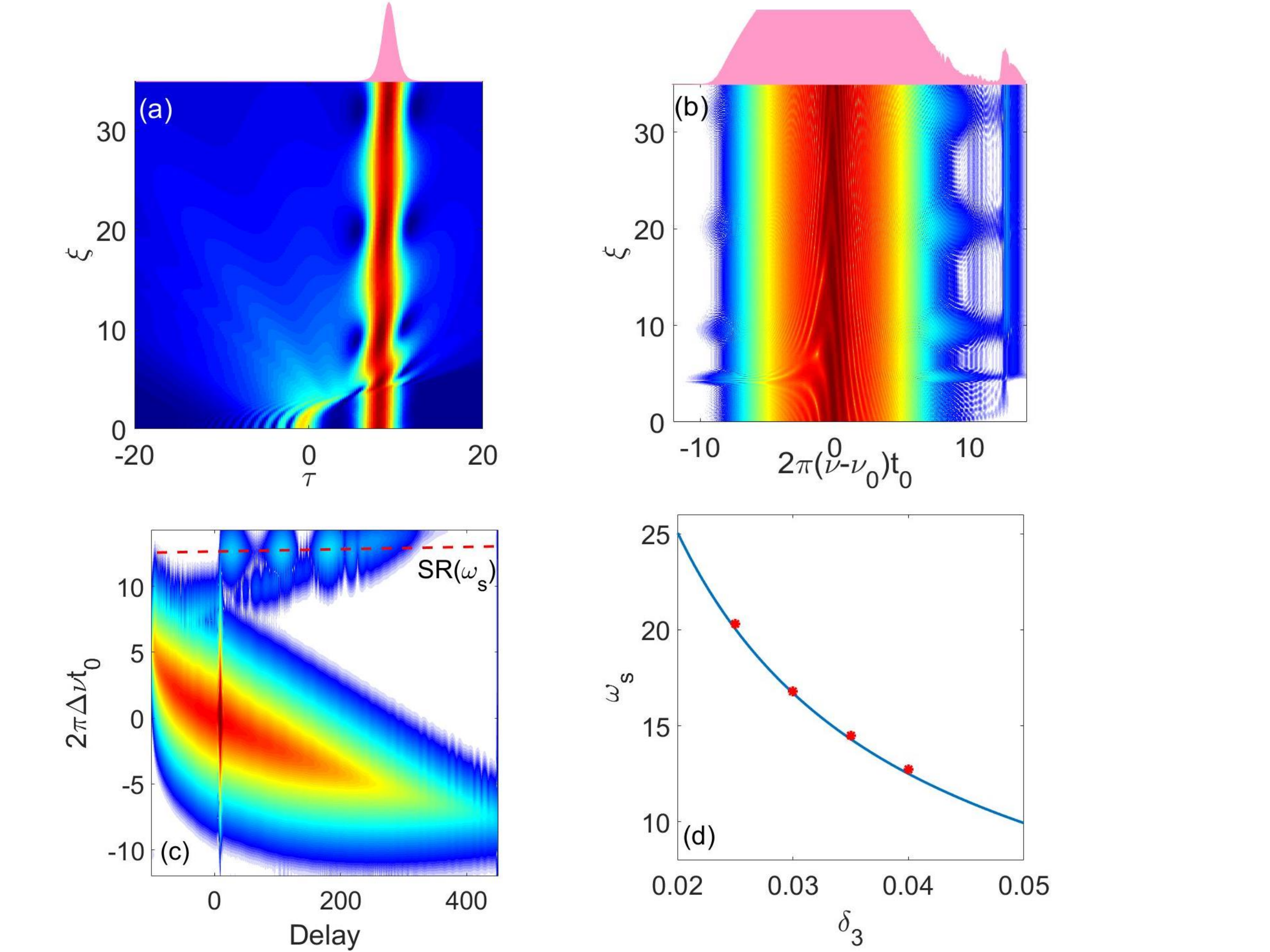}

           \caption{ The (a)temporal and (b) spectral dynamics of the system for a fixed value of $\delta_3=0.04$. The final temporal and spectral distributions are attached with the figures.(c) The XFROG spectrogram of the system at the final spatial point. the patch of the strong radiation is evident on the blue side of the spectrum.(d)The comparison of analytical values of $\omega_s$ (blue solid line) and numerically found (red dots) values with respect to $\delta_3$.  }
           \label{fig10}
           \end{center}
          
           \end{figure}

 The presence of positive TOD distorts the pulse shape significantly and Airy pulse experiences a temporal flipping at $\xi_{flip}$.    Hence the temporal separation ($\Delta \tau$) between the Airy pulse and the soliton cannot be too large at the launching position. However $\Delta \tau$ cannot be so close that their temporal distributions superimpose with each other and the pulses lose their individuality. Keeping this in mind, we have kept the value of $\Delta \tau$ at 8 and have plotted the dynamics in Fig.\ref{fig10}. The value of $\delta_3$ is also kept low for these simulations ($\delta_3$=0.04).  A large value of $\delta_3$ reduces the flipping point significantly and the Airy pulse flips before it hits the soliton. In the spectral plot (Fig.\ref{fig10}(b))) the weak and the strong radiation is evident. It is interesting to note that, the spatial location of the collision-mediated strong radiation for $\delta_3>0$ can be determined using the expression Eq.\ref{q7}. Only the numeric sign of the TOD parameter will change in this case and this is the reason why radiation flips from red to blue. 
 The signature of the strong radiation is further confirmed in the spectrogram in Fig.\ref{fig10}(c). The origination of collision mediated radiation($\omega_s$) follows the same physics as presented in the earlier section. Note that, the expression of $\omega_s$ in Eq.\ref{q12} is valid for this case also. In Fig.\ref{fig10}(d) we have compared the analytical expression(solid blue line) with the numerically found values(red dots) which  corroborate well.
 
 \subsubsection{Dynamics beyond the flipping point($\xi>\xi_{flip}$)} 

\begin{figure}[h!]
       \begin{center}
       \includegraphics[trim=5.0in 0.00in 6in 0in,clip=true,  width=42mm]{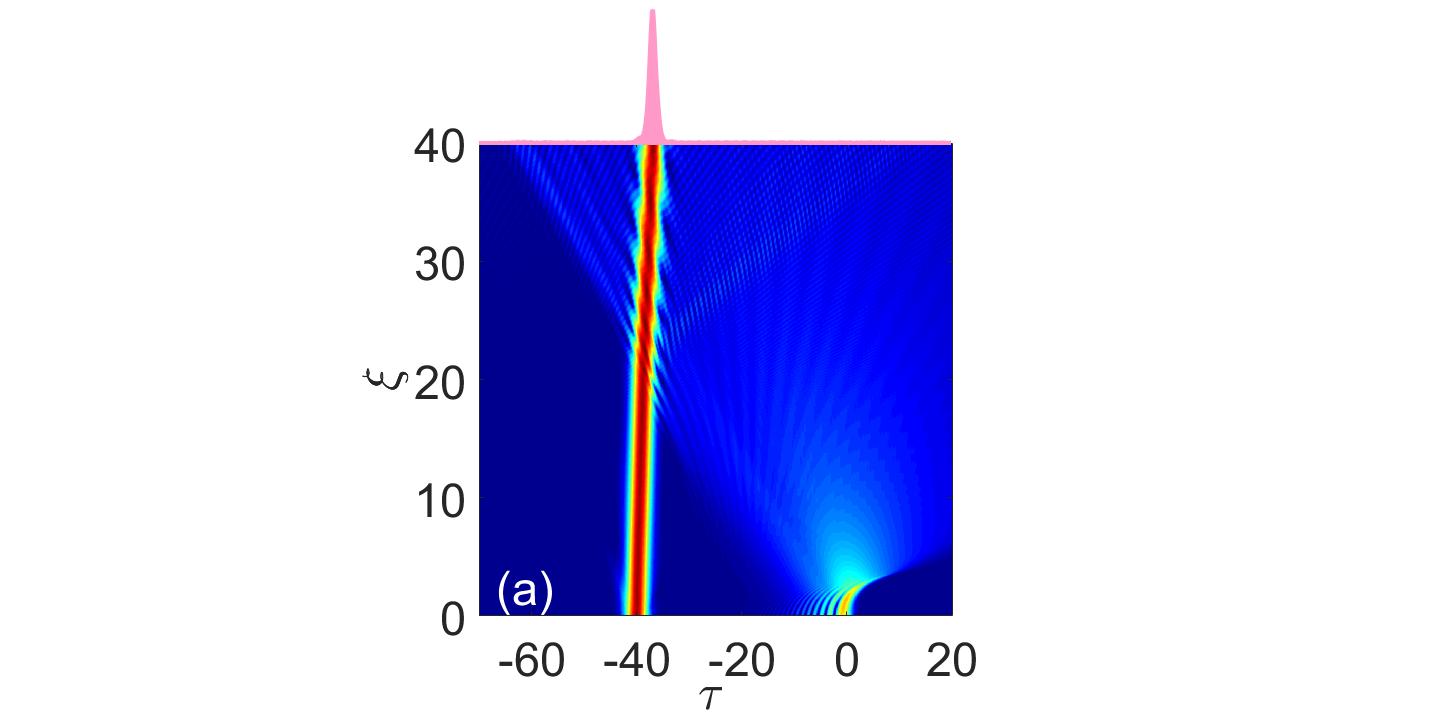}
      \includegraphics[trim=5in 0.00in 6in 0.0in,clip=true,  width=42mm]{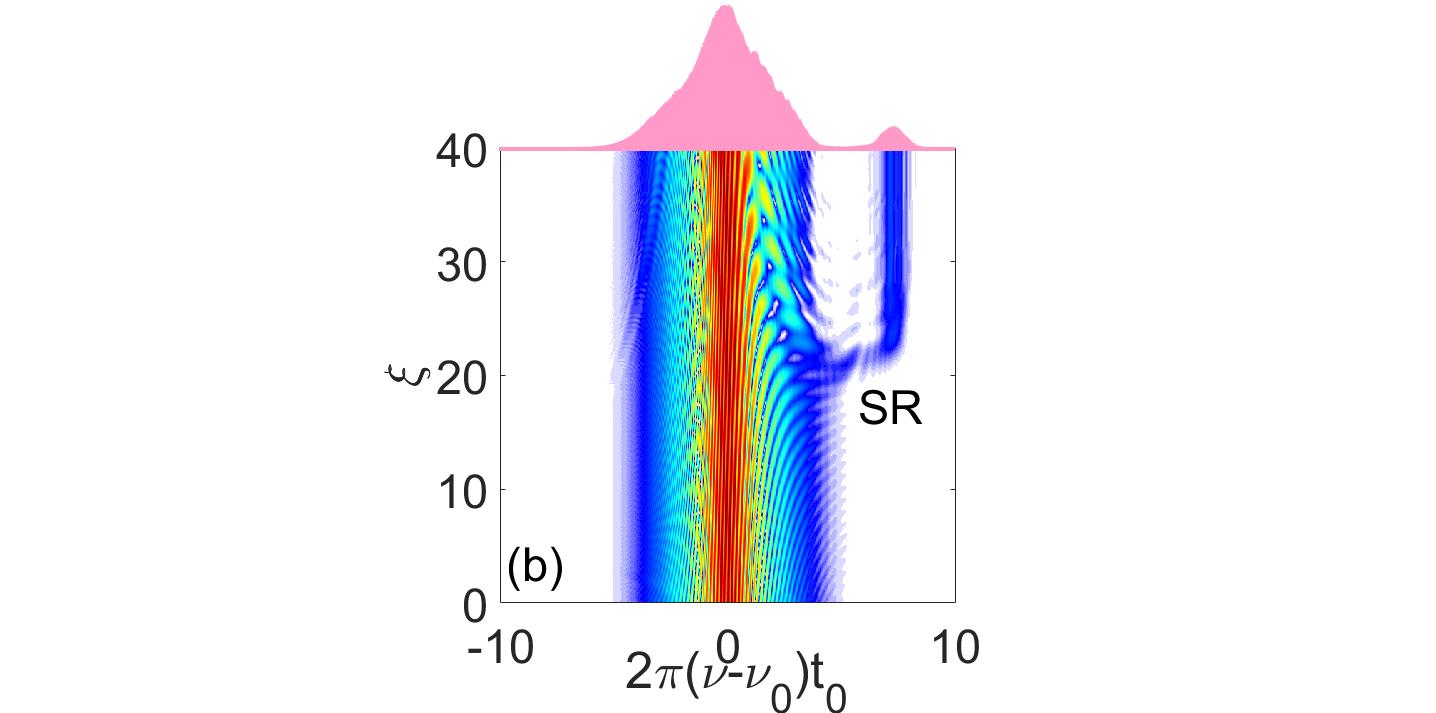}
      \includegraphics[trim=5.0in 0.00in 6in 0.0in,clip=true,  width=42mm]{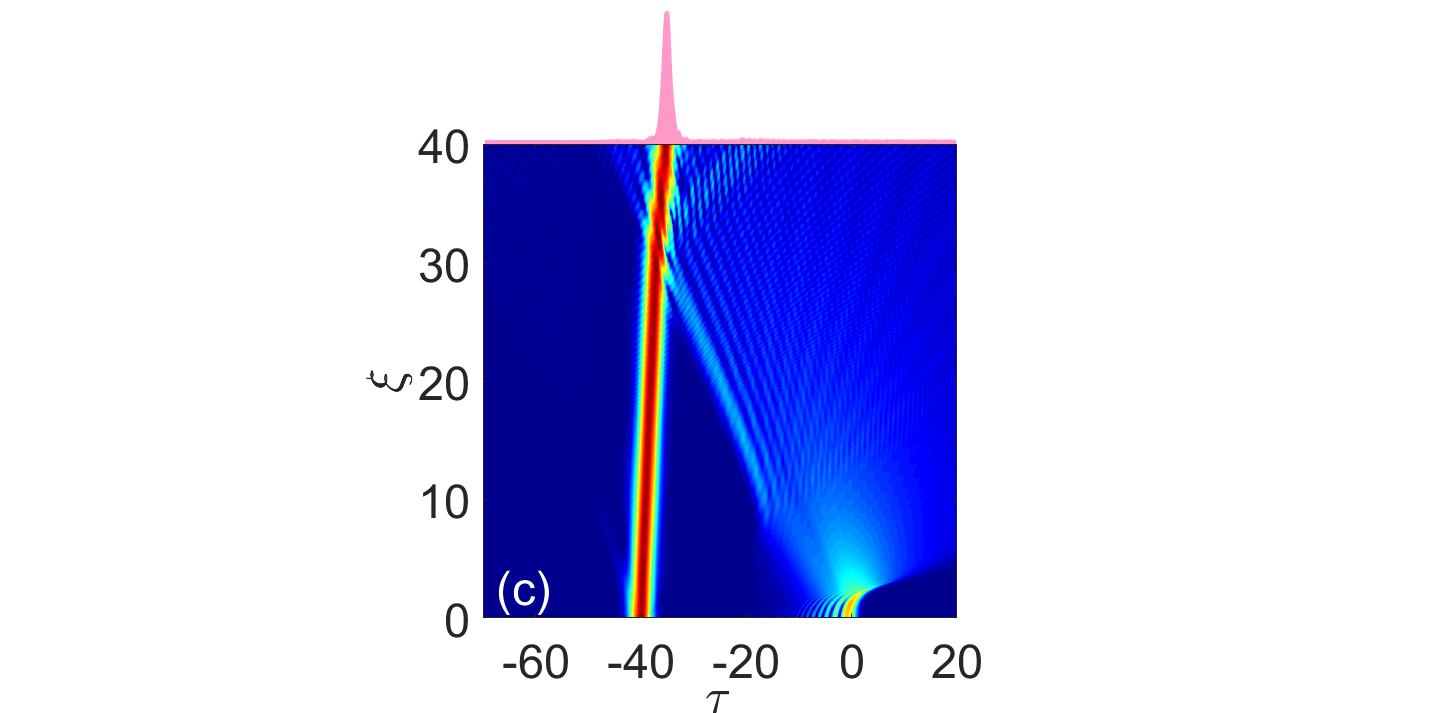}
      \includegraphics[trim=5.0in 0.00in 6in 0.0in,clip=true,  width=42mm]{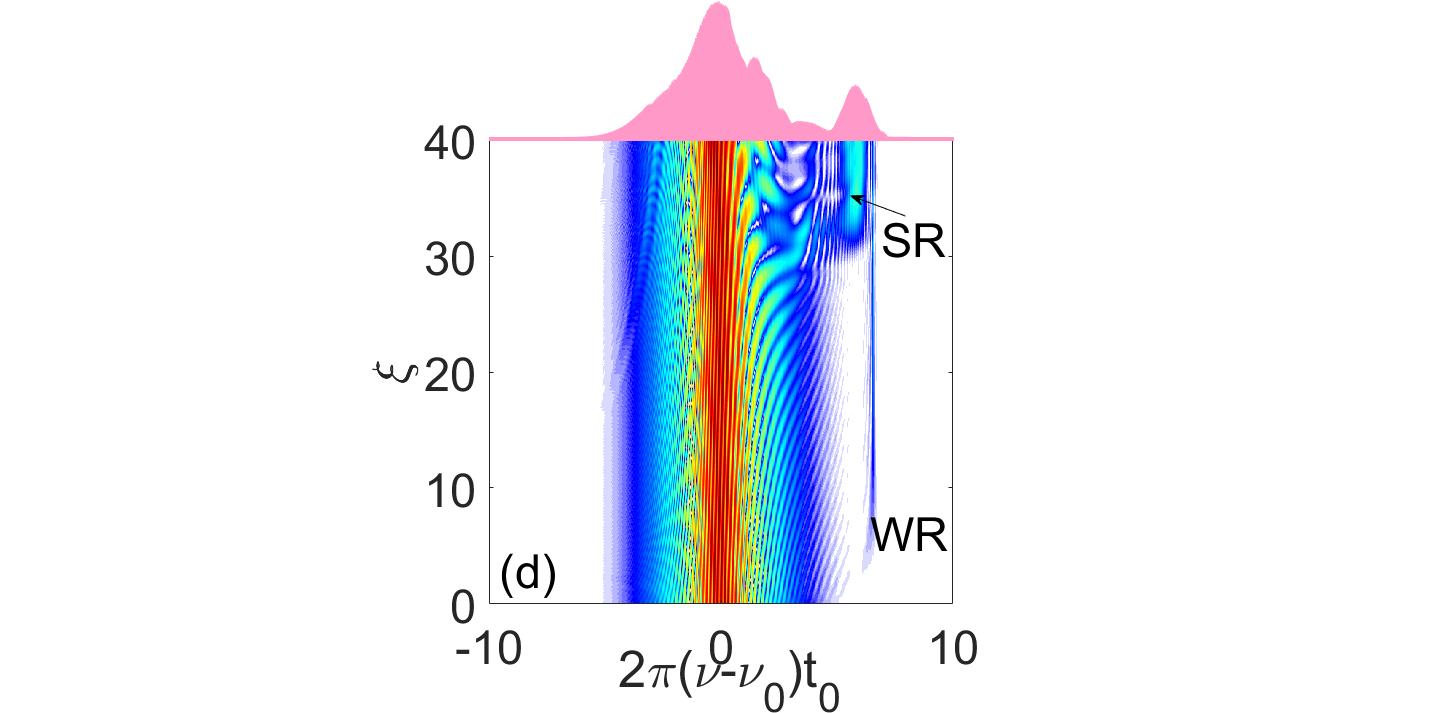}
      
       \caption{The dynamics of the pulses for different TOD parameters in temporal (0.06 for (a) and 0.08 for (c)) and in spectral domain 0.06 for (b) and 0.08 for (d)). The truncation parameter of the Airy pulse is $a$=0.1 and time separation of the pulses is 40 at the input. We can see that the position of collisions of two pulses is different for different TOD parameters.}
       \label{fig11}
       \end{center}
      
       \end{figure}

\begin{figure}[h!]
  \begin{center}
\includegraphics[trim=4.5in 0.0in 5.2in 0.7in,clip=true,  width=42mm]{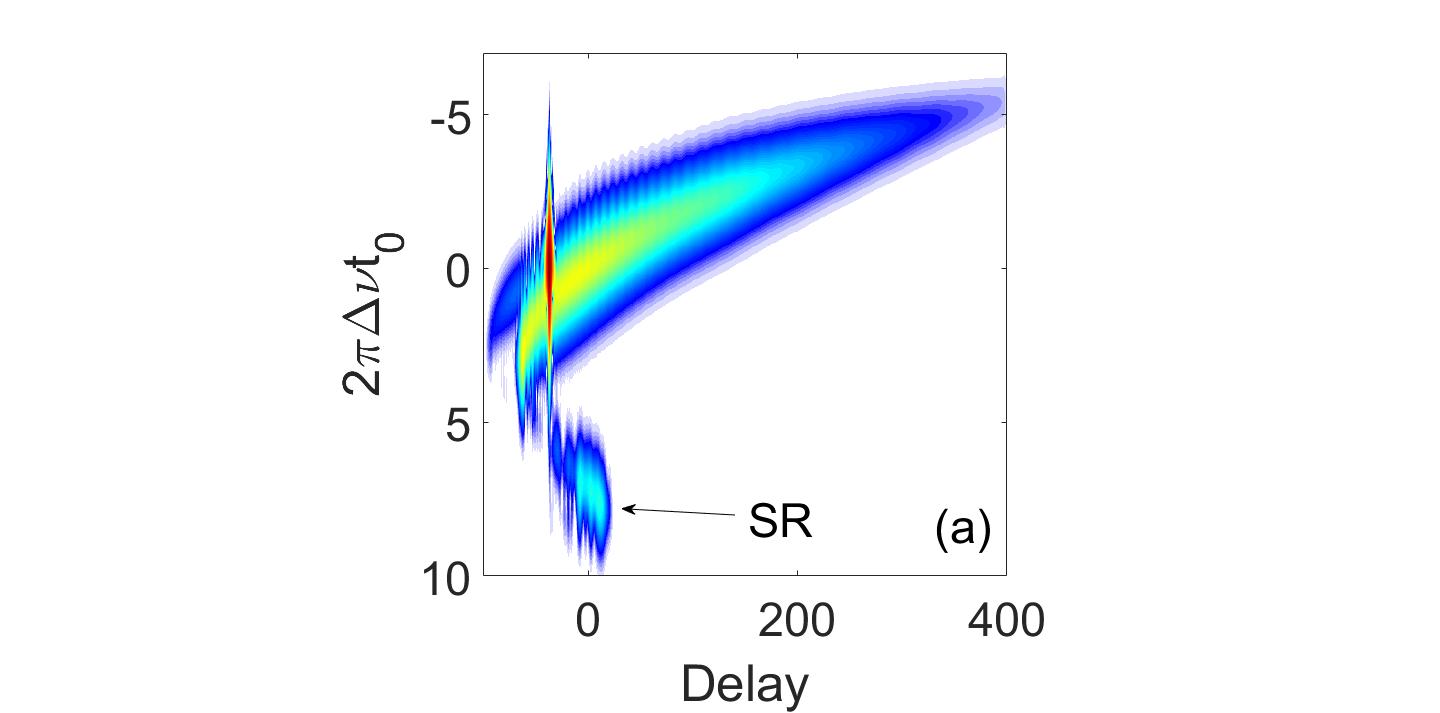}
\includegraphics[trim=4.5in 0.0in 5.2in 0.7in,clip=true,  width=42mm]{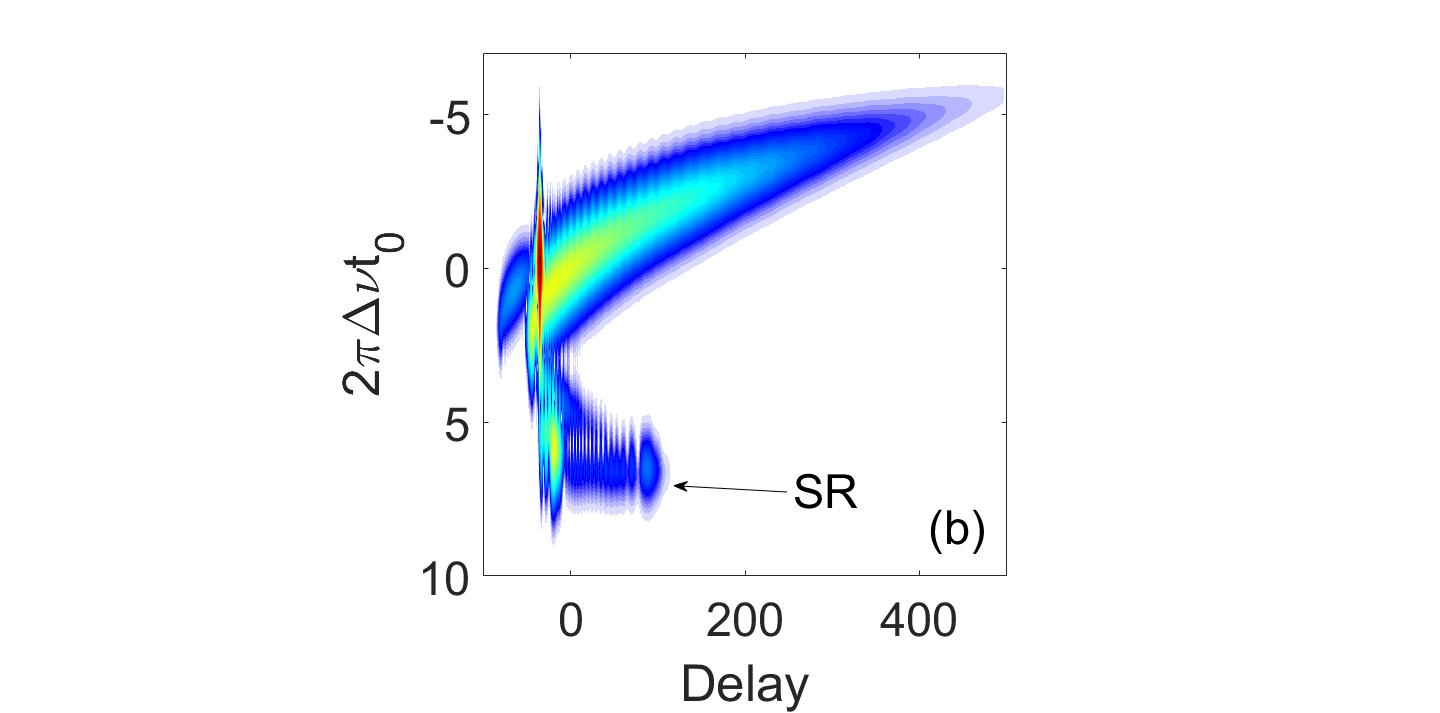}
\caption{The spectrograms of the pulses for different TOD parameters in  spectral domain($0.06$ for (a) and $0.08$ for (b)). The truncation parameter of the Airy pulse is $a=0.1$ and time separation of the pulses is $40$ at the input.The patch of SR is indicated by the arrows in the figures.}
\label{fig12}
\end{center}              
\end{figure}

In this section we theoretically investigate the  physics of the collision mediated radiation that takes place beyond the flipping point. In this case  the Airy pulse accelerates in the reverse direction and hits the soliton at $\xi>\xi_{flip}$. This phenomenon is rather intriguing and demands extensive study. The approach towards the problem is not trivial here as the Airy pulse flips at the singular point and changes its shape. A major challenge is to obtain the analytical expression of the Airy pulse beyond the flipping point ($\xi>\xi_{flip}$). It is obvious that, to meet the condition of temporal collision at $\xi>\xi_{flip}$, we launch the soliton in advance. In Fig.(\ref{fig11}) we show the collision dynamics for two different $\delta_3$ values. The spectrogram is also shown in Fig.(\ref{fig12}). The strong radiation is readily evident when the reversed Airy pulse hits the soliton. The phase of the Airy pulse can be calculated easily under negative TOD as the pulse doesn't face any singularity. But in case of the positive TOD the Airy pulse experiences a singularity in ($\xi$-$\tau$) space and the temporal distribution is inverted. At first stage, the Airy pulse is converted to a Gaussian pulse at  singularity. Then that Gaussian pulse leads to a new Airy pulse with inverted temporal wings. The form of the Gaussian pulse at singularity is given by \cite{Roy}, 

\begin{equation}\label{q13}
U(\xi_{flip},\tau)=U_0 \exp \left[-\frac{(\tau-a^2)^2}{\tau_{f}^2} \right]\exp(i\phi),
\end{equation}
where the amplitude ($U_0$) and phase ($\phi$) are,
\begin{equation}\label{q14}
U_0= \frac{1}{2\sqrt{\pi\gamma}}\exp(a^3/3), 
\end{equation}
and
\begin{equation}\label{q15}  
\phi=\frac{(\tau-a^2)^2}{24\delta_3\gamma^2}-\frac{1}{2}\tan^{-1}\left(\frac{1}{6a\delta_3}\right).   
\end{equation} 

   \begin{figure}[h!]
   \begin{center}
   \includegraphics[trim=4.3in 0.0in 4.5in 0.2in,clip=true,  width=42mm]{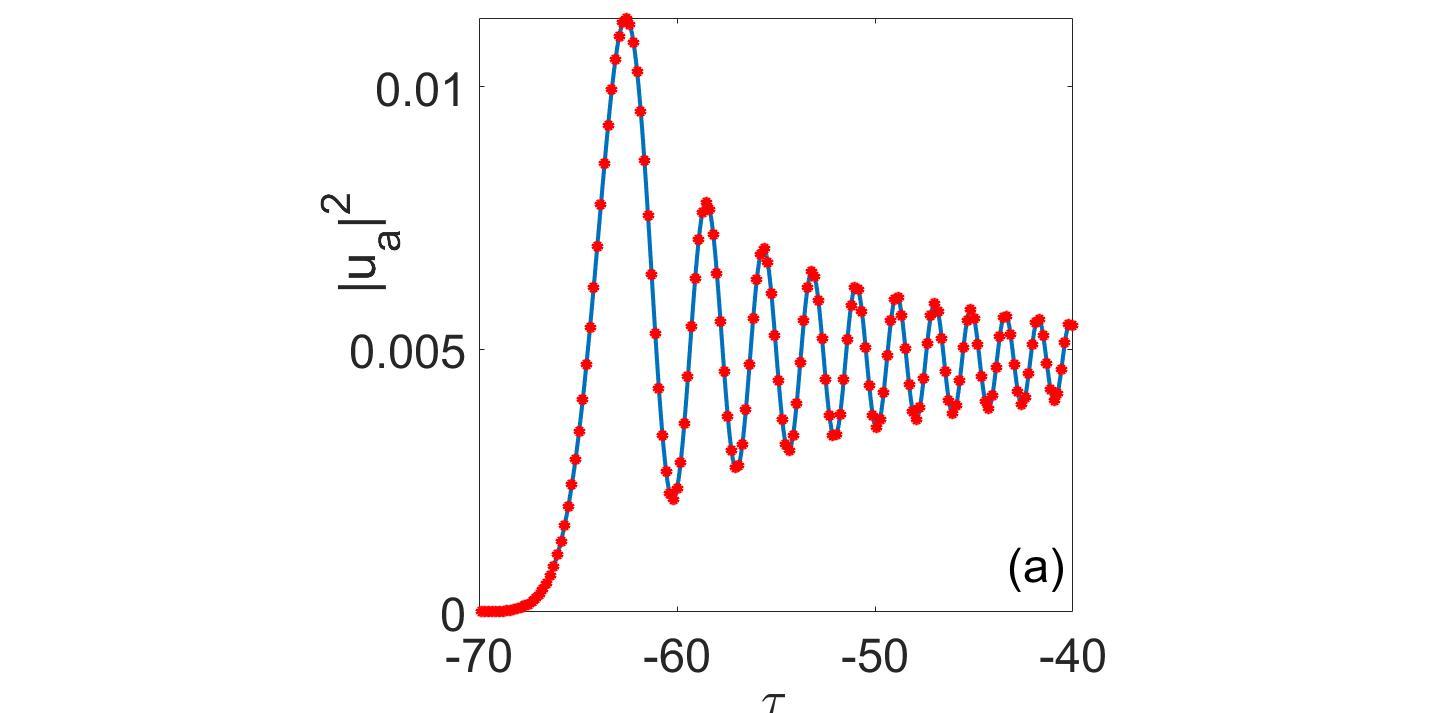}
   \includegraphics[trim=4.3in 0.0in 4.5in 0.2in,clip=true,  width=42mm]{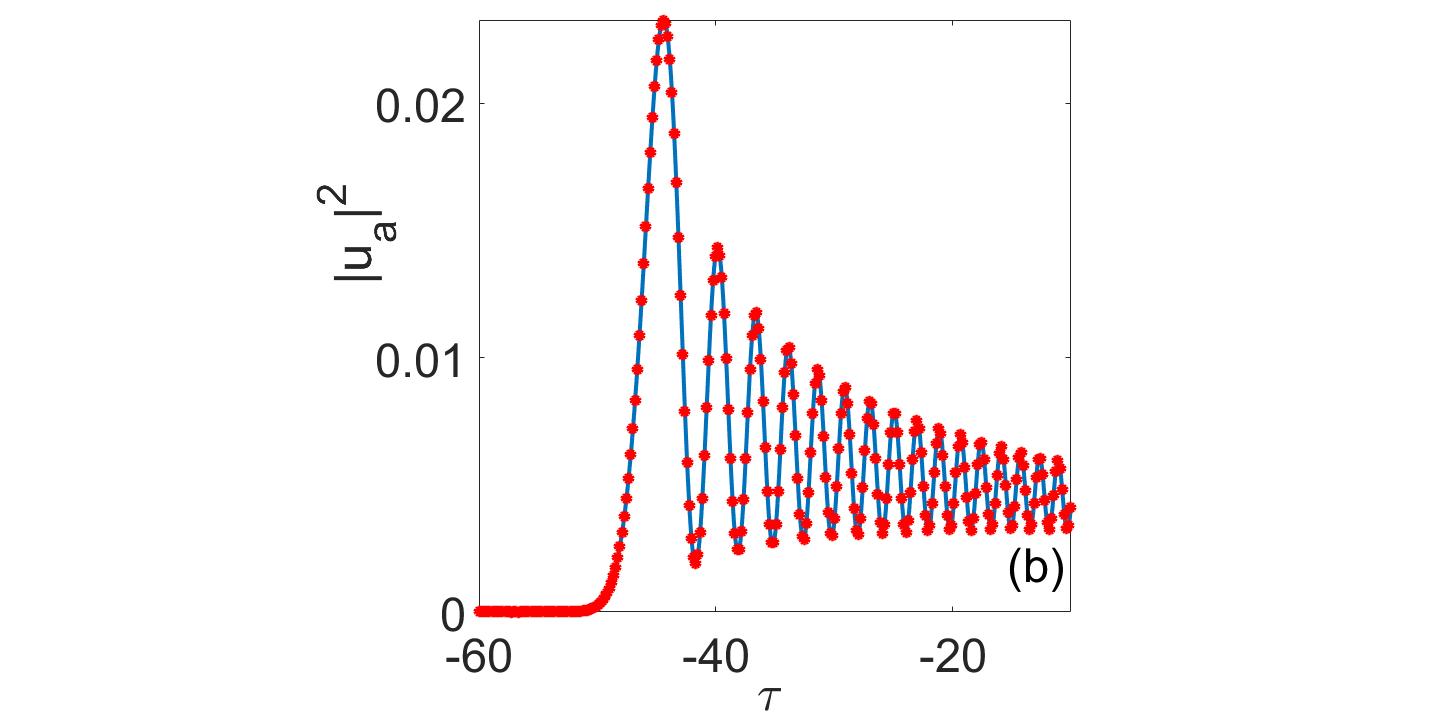}
   \caption{The comparison between the analytical solution (solid line) and the numerical solutions (dots) of the inverted Airy pulse in presence of positive third order dispersion with $\delta_3=0.06$ for (a) and $\delta_3=0.08$ for(b).The solutions are taken at a fixed distance ($\xi=40$)}
   \label{fig13}
    \end{center}
    \end{figure} 
The characteristic width of the Gaussian pulse is, $\tau_f=2\chi\sqrt{a}$ and  $\gamma =a\chi $ with $\chi=\sqrt{(1+1/36a^2\delta_3^2)}$. We consider this Gaussian pulse as our initial pulse to obtain the analytical form of the reversed Airy pulse at $\xi>\xi_{flip}$ . Rescaling the propagation distance as, $\xi'=(\xi-\xi_{flip})$, we finally obtain the  analytical expression of the Airy pulse at  $\xi>\xi_{flip}$,  
\begin{equation}\label{q16}
   u_{a}(\xi',\tau)=\frac{1}{c'} \exp \left(\frac{a^3}{3}\right)Ai\left(\frac{b'}{c'}-\frac{n'^2}{c'^4}\right)\exp i \left( \frac{2n'^3}{3c'^6}-\frac{n'b'}{c'^3} \right) 
   \end{equation}
The parameters are defined as  $c'=(3\delta_3\xi')^{\frac{1}{3}}$;  $n'=(ia+sgn(\beta_2)\xi'/2-1/6\delta_3);$ and $b'=-\tau $.  Note that, except the scaling factor, the expression of reversed Airy pulse is exactly same that we obtained in Eq. \eqref{q4}. To check the validity of the expression given in Eq. \eqref{q16}, we compare the analytical expression with the numerical simulations as shown in  Fig.\ref{fig13}(a,b). The phase of the reversed Airy pulse will have the form, $\phi_a'=\Gamma/c'^6$, where $c'=(3\delta_3\xi')^{1/3}$. The inverted pulse beyond the flipping point follows a non-parabolic trajectory. The expression of the temporal position ($\tau_p'$) in ($\xi$-$\tau$) space is given as,
   \begin{equation}\label{q17}
   \tau_{a}'\approx-\frac{(\frac{\xi'}{2}+\frac{1}{6\delta_3})^2}{3\delta_3\xi'}
   \end{equation}
   
 The phase of the reversed Airy pulse at the collision point $\xi'_c$ is 
   $\phi'_{a}\approx \xi_c^{3}/{24c_c'^6}$ where $c_c'=(3\delta_3\xi'_c)^\frac{1}{3}$. Exploiting the phase of the reversed Airy pulse we obtain the radiation frequency as , 
   \begin{equation}\label{q18}
   \omega_s=\frac{1}{2\delta_3}+4\delta_3-4\delta_3\frac{\phi'_{a}}{\xi'}\bigg\vert_{\xi'=\xi'_c}
   \end{equation}

\begin{figure}[]
       \begin{center}
       \includegraphics[trim=4.5in 0.09in 5.55in .5in,clip=true,  width=42mm]{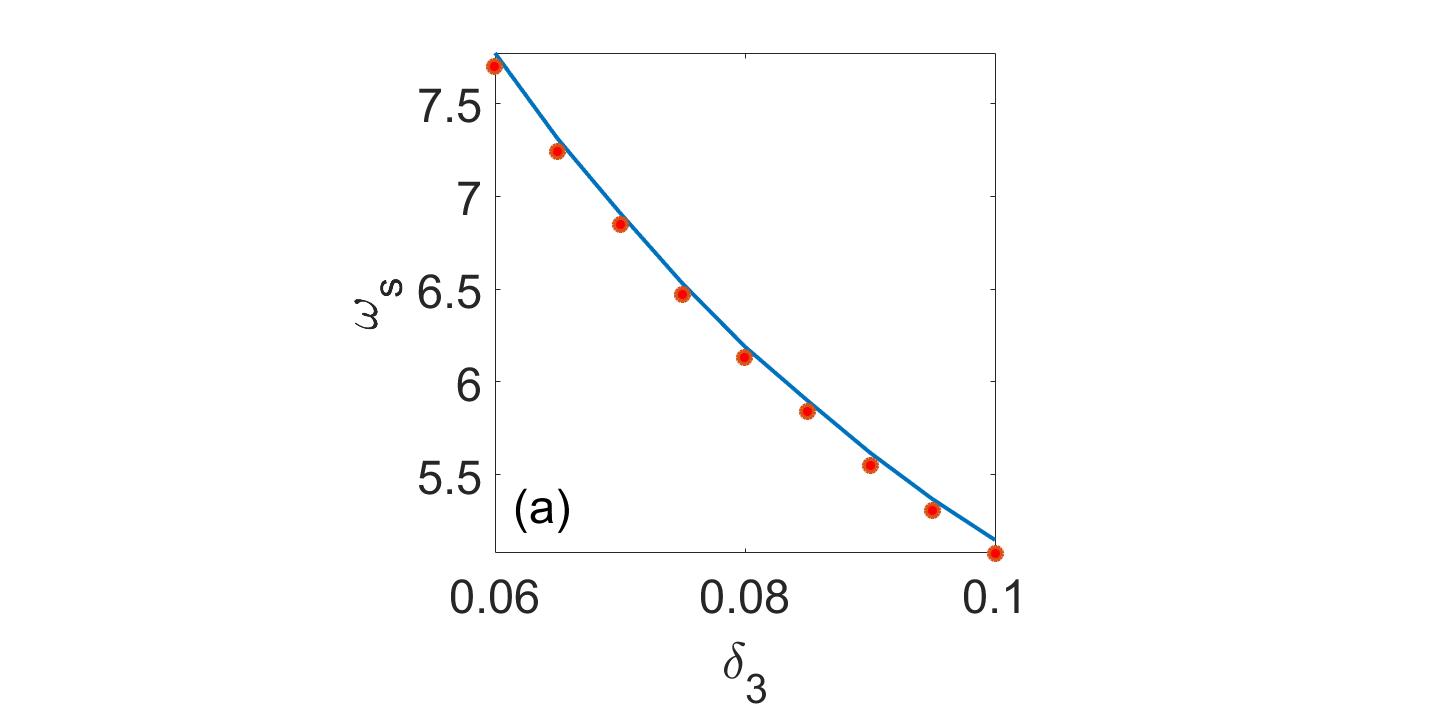}
       \includegraphics[trim=4.5in 0.09in 5.55in .5in,clip=true,  width=42mm]{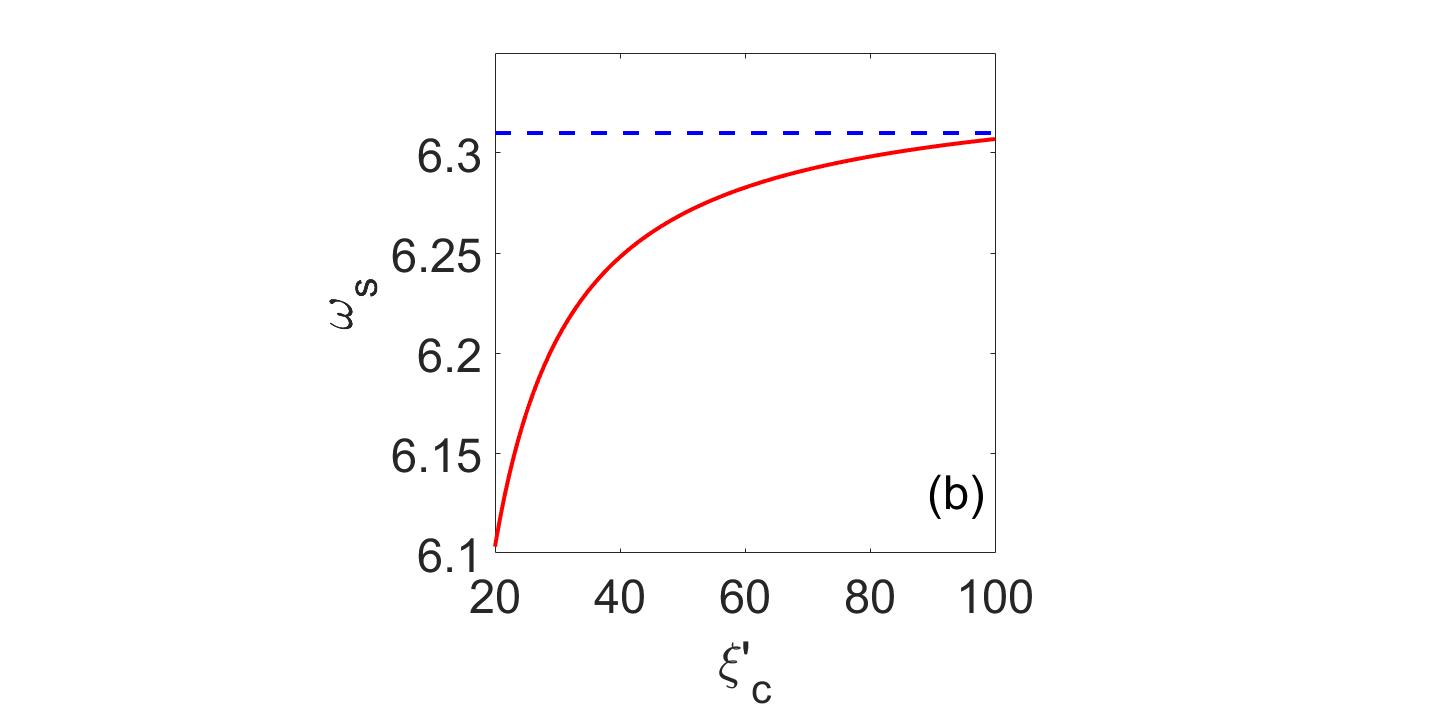}
        \caption {(a) Variation of $\omega_s$ with TOD parameter $\delta_3$. The theoretical result (solid blue line) agrees well with numerical data (dots). (b) $\omega_s$ as a function of $\xi'_c$. $\omega_s$ is hardly affected by point of collision.}
       \label{fig14}
       \end{center}
       \end{figure} 
 In Fig.\ref{fig14}(a) we plot the SR frequency ($\omega_s$) as a function of TOD coefficient $\delta_3$. The solid line indicates the analytical expression that we obtain from Eq. (\ref{q18}), where dots are the radiation frequencies measured numerically. In Fig.\ref{fig14}(b) we have plotted  the analytical expression of $\omega_s$  as a function of $\xi'_c$.  It is evident from the figure that in this case also the value of $\omega_s$ does not vary significantly  and after a certain distance it saturates to a value (indicated by the blue dashed line in the Fig.\ref{fig14}(b)) which is close to the value of WR frequency.

\section{Conclusion}

In this report we have studied the interaction of co-propagating Airy pulse and soliton  in a Kerr medium under TOD.  The power of the Airy pulse is controlled in such a way that it doesn't shed any soliton and retains its characteristic shape. It is observed that, new spectral component is emerged in the form of radiation when the self-accelerating finite energy Airy pulse collides with a delayed soliton. We emphasis that, this collision-mediated radiation is characteristically different from the well-known Cherenkov radiation emitted by a perturbed soliton and hence demands special analytical treatment. Interestingly, this collision-assisted radiation takes place only in the environment of TOD and sensitive to the numeric sign of TOD parameter.  To realise the system physically we proposed a realistic waveguide structure based on Si which exhibits both positive and negative TOD at two different operational wavelength. We calculate the values of dispersion and nonlinear parameters for the proposed waveguide and use them in our numerical simulation. The dynamics of the Airy pulse is dramatic under positive TOD where it experiences a singularity and flips in time domain. Depending on the numeric sign of TOD parameter we judiciously launched the soliton so that collision takes place with accelerating (or decelerating) Airy pulse.   We develop a theory to explain the collision-mediated radiation for both positive and negative TOD and derive the frequency of the detuned radiation. We compare our findings with numerical results and our theoretical predictions are found to be in well agreement with the direct numerical modelling of the  generalised nonlinear Schr\"{o}dinger equation (GNLSE).         
\section*{Acknowledgment}

A. B. acknowledges MHRD, India for his research fellowship.

\end{document}